\documentclass[sigconf, nonacm]{acmart}%

\usepackage{balance}    %
\usepackage{color}
\usepackage{color, colortbl}
\usepackage{array}
  
\usepackage{caption}
\usepackage{subcaption}

\usepackage{multirow}

\usepackage{verbatim}  %

\usepackage{xspace}

\usepackage{graphicx}
\usepackage{xcolor}

\definecolor{Gray}{gray}{0.97}
\definecolor{MedGray}{gray}{0.9}
\definecolor{greytext}{gray}{0.5}
\definecolor{DarkGreen}{rgb}{0.0, 0.5, 0.0}
\definecolor{PFGreen}{rgb}{0.0, 0.5, 0.0}
\definecolor{lightGreen}{rgb}{0.8, 0.9, 0.8}
\definecolor{CadmiumGreen}{rgb}{0.0, 0.42, 0.24}
\definecolor{DarkKhaki}{rgb}{0.74, 0.72, 0.42}
\definecolor{DarkRed}{rgb}{0.7, 0.2, 0.2}
\definecolor{Purple}{rgb}{0.7,0.0,0.7}
\definecolor{Brown}{rgb}{0.7,0.3,0}
\definecolor{Orange}{rgb}{1, 0.5, 0.1}
\definecolor{niceblue}{rgb}{0.0, 0.2, 0.4}

\newcommand{\pcbrenewal}[0]{\textsc{PCB Renewal}\xspace}

\graphicspath{
{figures/} %
}

\usepackage[nameinlink,capitalise]{cleveref}
\crefname{enumi}{}{}  %
\crefrangeformat{figure}{Figures~#3#1#4--#5#2#6}

\usepackage{siunitx}
\usepackage{algorithm2e}
\usepackage{amsmath}
\AtBeginDocument{%
  \providecommand\BibTeX{{%
    \normalfont B\kern-0.5em{\scshape i\kern-0.25em b}\kern-0.8em\TeX}}}

\setcopyright{none}
\renewcommand\footnotetextcopyrightpermission[1]{} %
\begin{document}

\title[]{\pcbrenewal: Iterative Reuse of PCB Substrates for \\ Sustainable Electronic Making}
\author{Zeyu Yan}
\email{zeyuy@umd.edu}
\affiliation{%
  \institution{University of Maryland}
  \city{College Park}
  \state{Maryland}
  \country{USA}
  \postcode{20895}
}

\author{Advait Vartak}
\email{avartak@terpmail.umd.edu}
\affiliation{%
  \institution{University of Maryland}
  \city{College Park}
  \state{Maryland}
  \country{USA}
  \postcode{20895}
}

\author{Jiasheng Li}
\email{jsli@umd.edu}
\affiliation{%
  \institution{University of Maryland}
  \city{College Park}
  \state{Maryland}
  \country{USA}
  \postcode{20895}
}

\author{Zining Zhang}
\email{znzhang@umd.edu}
\affiliation{%
  \institution{University of Maryland}
  \city{College Park}
  \state{Maryland}
  \country{USA}
  \postcode{20895}
}

\author{Huaishu Peng}
\email{huaishu@umd.edu}
\affiliation{%
  \institution{University of Maryland}
  \city{College Park}
  \state{Maryland}
  \country{USA}
  \postcode{20895}
}

\renewcommand{\shortauthors}{Yan, et al.}

\begin{abstract}

PCB (printed circuit board) substrates are often single-use, leading to material waste in electronics making. We introduce \pcbrenewal, a novel technique that ``erases'' and ``reconfigures'' PCB traces by selectively depositing conductive epoxy onto outdated areas, transforming isolated paths into conductive planes that support new traces. We present the \pcbrenewal workflow, evaluate its electrical performance and mechanical durability, and model its sustainability impact, including material usage, cost, energy consumption, and time savings. We develop a software plug-in that guides epoxy deposition, generates updated PCB profiles, and calculates resource usage. To demonstrate \pcbrenewal’s effectiveness and versatility, we repurpose  a single PCB across four design iterations spanning three projects: a camera roller, a WiFi radio, and an ESPboy game console. We also show how an outsourced double-layer PCB can be reconfigured, transforming it from an LED watch to an interactive cat toy. The paper concludes with limitations and future directions.

\end{abstract}

\ccsdesc[500]{Social and professional topics~Sustainability}
\ccsdesc[500]{Hardware~Printed circuit boards}
\ccsdesc[300]{Human-centered computing~Systems and tools for interaction design}

\keywords{PCB Prototyping, Sustainability, Reuse, Renewal,  Fabrication}

\begin{teaserfigure}
  \includegraphics[width=\textwidth]{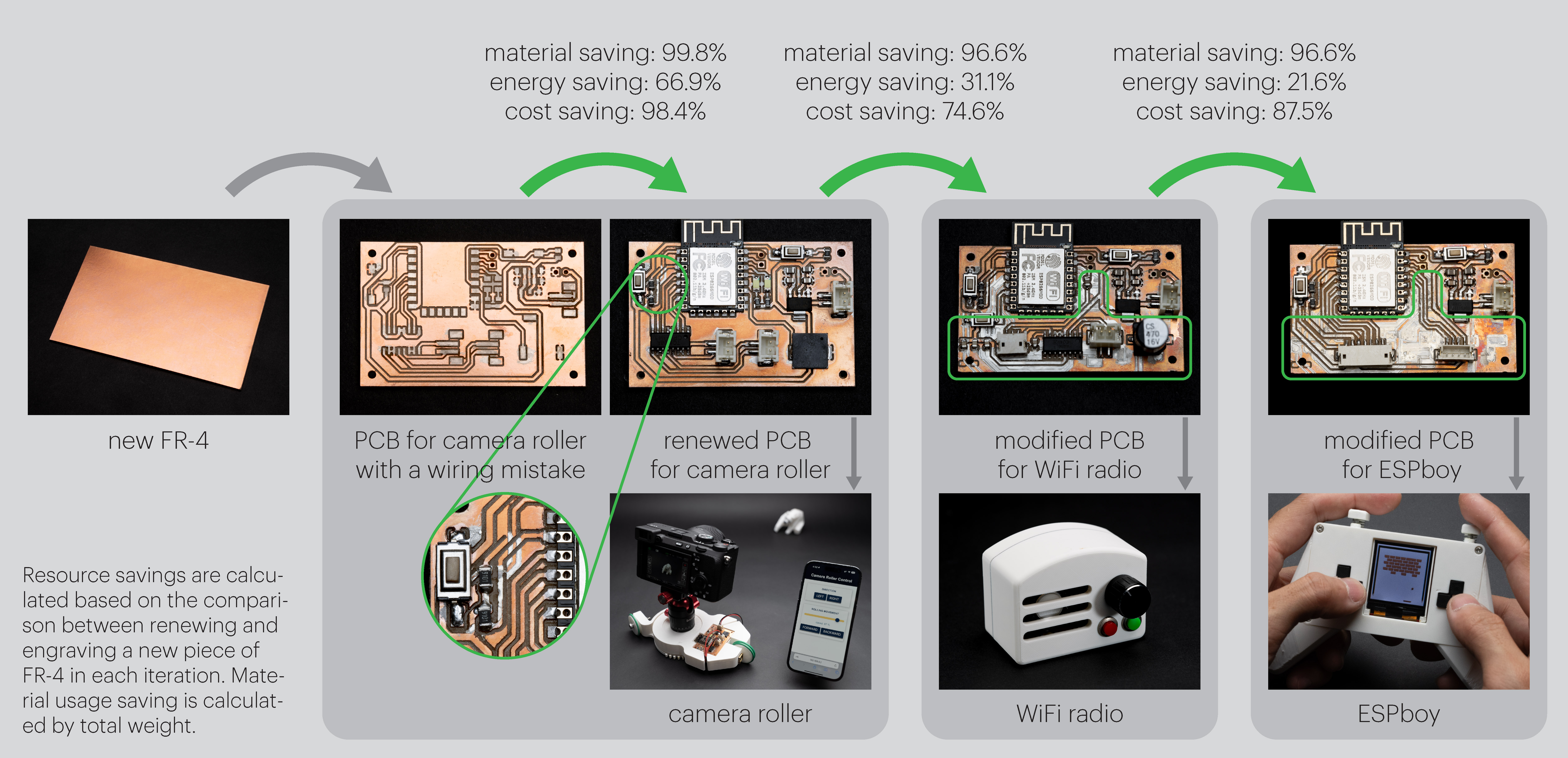}
  \caption{\pcbrenewal: a single piece of FR-4 was renewed for four iterations across three projects, significantly reducing resource consumption compared to engraving new circuits for each iteration.}
  \label{fig:teaser}
  \Description{PCB Renewal: a single piece of FR-4 was renewed for four iterations across three projects, significantly reducing resource consumption compared to engraving new circuits for each iteration.}
\end{teaserfigure}

\maketitle

\section{Introduction}\label{Intro}
Printed circuit boards (PCBs) are critical components in nearly all electronic devices. However, the obsolescence of PCBs, from their design process to end-of-life disposal, has become an increasingly significant source of electronic waste (e-waste).

The design of a functional PCB typically involves multiple stages, including software simulation, circuit validation (e.g., via breadboarding), and prototyping with custom PCB batches. While the simulation and breadboard validation phases generate minimal e-waste—since engineers test functionality digitally or reconfigure reusable components like breakout boards, through-hole electronics, and jumper wires—the subsequent PCB prototyping stage inevitably contributes to e-waste production.

PCBs are made using a subtractive fabrication method, where copper layers are permanently etched from laminated substrates (e.g., FR-4, a fiberglass-reinforced epoxy), making the process \textit{inherently irreversible}. During PCB prototyping, minor errors—such as flaws in electronic design automation (EDA) schematics or mismatches between PCB dimensions and their housing—are often discovered. While these issues may be small and easily corrected digitally, they are physically embedded into the substrates, rendering the entire prototype (or batches, if outsourced to factories) unusable. This necessitates the repeated production of new PCBs, while discarded ones contribute to e-waste.

Mass-produced PCBs further exacerbate the e-waste problem when devices reach the end of their life cycle. In 2022, less than 23\% of globally generated e-waste was formally collected and recycled. Even when PCBs are recycled, their inherently irreversible fabrication process forces them into centralized waste streams, where they are processed indiscriminately. As a result, they are rarely repaired, repurposed, or reused—even though many PCBs and their substrates remain functional~\cite{e-waste}.

These e-waste challenges have garnered attention in the HCI community, as evidenced by sustainable making and unmaking workshops at UIST and CHI~\cite{yan2023future, 10.1145/3613905.3636300}, and a dedicated TOCHI special issue~\cite{song2025unmaking}.
Recent work has also called for a reimagining of end-users' roles, emphasizing their potential not only as consumers but also as active participants in PCB recycling and reuse~\cite{lu2024unmaking}. Additionally, researchers have advocated for the development of new processes, tools, and infrastructure to address e-waste and promote sustainable practices~\cite{MakeMaking}.

In this paper, we contribute to sustainable PCB practices by proposing a \textit{reversible} PCB substrate fabrication process that enables the ``erasure'' and ``reconfiguration'' of copper layouts.
Central to this process is the additive restoration of removed copper areas using conductive fillers, such as conductive epoxy, to renew the PCB substrate for fresh trace patterns. 
Analogous to a correction pen overwriting mistakes on paper, our approach extends the lifespan of PCB substrates by enabling physical re-editing to correct design errors or remove obsolete traces. This transforms what would otherwise become e-waste into new designs (Figure~\ref{fig:teaser}). We call this approach \pcbrenewal.

In the remainder of this paper, we introduce the workflow of \pcbrenewal, providing a detailed examination of conductive filler materials and the key fabrication processes involved in the renewal of the commonly used PCB substrate FR-4.
We validate our approach through a series of experiments that evaluate key electrical parameters, including conductivity, current capacity, solder joint durability, and the number of renewal iterations a single FR-4 board can undergo.
These experiments demonstrate that the renewed substrate exhibits electrical performance comparable to that of raw FR-4.
To assess the sustainability impact of \pcbrenewal, we present a quantitative analysis model that compares \pcbrenewal with the fabrication of new circuits using raw FR-4. 
This model includes estimates of material usage, cost, time, and energy consumption.
To help end-users incorporate \pcbrenewal into their workflow to save PCB substrates during prototyping or repurpose PCB designs in general, we develop an EDA software plug-in. 
This plug-in allows end users to update a circuit design with changes visualized across iterations, evaluate the sustainability impact of specific renewed designs, and generate the fabrication profiles required for renewal.  

\pcbrenewal can be applied to PCBs fabricated either in-house or through outsourcing. 
To demonstrate its versatility, we provide a detailed account of a single PCB reused across four in-house design iterations for three distinct projects: a wireless camera roller, a WiFi radio, and an ESPboy game console. 
Additionally, we demonstrate that an outsourced double-layer PCB, originally made for an LED watch, can be renewed and repurposed for a cat toy using the \pcbrenewal process. 
We report the sustainability impact of each design iteration for all examples. We conclude with a discussion on the limitations of \pcbrenewal and its potential future directions.

\section{Related Work}
Our work is inspired by a substantial body of prior research in sustainable human-computer interaction (SHCI), methods for recycling or reusing electronic and electronic waste, as well as technical explorations in PCB substrate repair and renewal.

\subsection{Sustainability in HCI: Making and Prototyping}

The notion of Sustainable Interaction Design (SID) was introduced by Blevis~\cite{10.1145/1240624.1240705} over a decade ago, providing a foundational framework for addressing environmental impacts and human behavior in the design of interactive technologies. 
This concept has since evolved into the broader field of SHCI.

Early discussions in SHCI often centered on mobile applications and their influence on end-users' daily behaviors, such as reducing energy consumption through persuasive computing~\cite{froehlich2009ubigreen, fogg2002persuasive}. 
More recently, attention has shifted to the environmental impact of making and physical prototyping~\cite{yan2023future, 10.1145/3613905.3636300}, driven by the democratization of personal fabrication tools and the growing maker movement~\cite{hudson2016understanding, shewbridge2014everyday}.

Several studies have explored end-users' (creative) approaches to engaging with wasted physical materials in daily activities. 
For example, Yan et al.~\cite{MakeMaking} have presented a qualitative research that maps out the sustainability practices, challenges and opportunities in modern makerspace setups and have called for new tools and infrastructure to support making sustainably. Kim and Paulos~\cite{10.1145/1978942.1979292} have proposed a reuse composition framework, based on online surveys and observations, to inspire the creative reuse of material waste. 
Dew and Rosner~\cite{10.1145/3322276.3322320} have conducted design explorations that examine how designers conceptualize, manage, and rework waste materials in educational makerspaces. 
Similarly, Maestri and Wakkary~\cite{maestri2011understanding} have studied the intersection of repair and creativity within household settings. 
These ideas have since evolved into broader concepts, such as unmaking~\cite{10.1145/3411764.3445529}, uncrafting~\cite{murer2015crafting}, and unfabricating~\cite{10.1145/3313831.3376227}, which employ speculative or participatory design lenses to explore the afterlife of objects and materials.

Alongside the exploration of reusing daily waste, HCI researchers have begun investigating the use of decomposable and biodegradable materials in making. 
For example, several projects have proposed using edible materials~\cite{Backyard-Degradable_Wireless_Heating_Interface, Printable_Play-Dough} or substances derived from food waste~\cite{Coffee_Grounds} as construction materials for molding and 3D printing. 
Microbe-based materials, such as yeast~\cite{SCOBY} and fungi~\cite{Mycelium-based_Materials, Myco-Accessories}, as well as biomaterials derived from living organisms, including algae~\cite{Alganyl} and cellulose-based fibers~\cite{Cellulose-Based_Optical_Textile_Sensors, Bioplastics}, have also been proposed as building materials for the prototyping of interactive devices. 

In addition, new fabrication processes and tools have been developed to support more sustainable making practices. 
For example, Filament Wiring~\cite{deshpandeunmake} and Substiports~\cite{Substiports} introduce alternative fabrication pipelines that repurpose wasted 3D printing filament or failed prints for new designs. EcoThreads~\cite{ecothread} and Desktop Biofibers Spinning~\cite{spinning} have developed new machines and processes to make water-dissolvable yarns easily accessible for sustainable textile applications. 

Our work is greatly inspired by the aforementioned advancements in sustainable making, with a specific focus on the processes involved in PCB making. As discussed in the introduction, PCBs are among the largest contributors to e-waste. Our work aims to reduce this environmental impact.

\subsection{Supporting the Reuse and Recycling of Electronics}
E-waste recycling requires interdisciplinary research and collaborative practices.

In the electronics management industry, the primarily focus is on infrastructure and large-scale processes that can extract raw materials from PCB scrap. 
For example, chemical and mechanical techniques are used to recover valuable materials, including refractory metals and elements of the platinum group found in standard PCB waste~\cite{C8GC03688H, su131810357}. 
Although effective, these industrial and centralized approaches void the opportunities for PCBs that might be repurposed, repaired, or reused, and they may fall short as more individuals become involved in creating electronics through the democratization of making tools. 

Recent HCI literature points out that many end users are no longer just consumers of physical artifacts but also their creators. 
Consequently, they bear greater responsibility for managing the material waste generated during the individual making process~\cite{MakeMaking, lu2024unmaking}. 
In this context, much of the HCI research focuses on promoting the reuse and recycling of electronics at the individual level. For example, the CurveBoards project~\cite{10.1145/3313831.3376617} proposes a custom-shaped breadboard design that is versatile for rapid prototyping with form-specific requirements. 
CircuitGlue~\cite{lambrichts2023circuitgiue} reduces waste in prototyping by allowing easy integration and reuse of off-the-shelf components.
SolderlessPCB~\cite{10.1145/3613904.3642765} demonstrates a pressure-based PCB assembly method using 3D printed or CNC-made housings, allowing easy disassembly and reuse of surface-mounted components.
ecoEDA~\cite{10.1145/3586183.3606745} shows how interactive circuit design software, by integrating early-stage suggestions for utilizing recyclable electronic components from stock PCBs, can facilitate the reuse of electronics throughout the design process.

New, more environmentally friendly PCB materials and compositions have also been explored. For example, transesterification vitrimers have been proposed as PCB substrate materials, which can be recycled through polymer swelling, achieving a 98\% polymer recovery~\cite{zhang2024recyclable}. Several studies have investigated PCB substrates based on paper~\cite{10.1145/3381013, 10.1145/2493432.2493486, 10.1145/3173574.3174143}, wood~\cite{10.1145/3474349.3480191}, and water-soluble materials~\cite{bharath2020novel, 10.1145/3491101.3519823, 10.1145/3161165}. 
Water-soluble materials are particularly interesting in the context of sustainable electronics, as their degrading processes are controllable. 
This enables the creation of transient electronic prototypes~\cite{https://doi.org/10.1002/adfm.201301847, https://doi.org/10.1002/adma.201403164} with programmable lifespans, simplifying the recycling of materials once they are no longer needed~\cite{cheng2023functional, song2023vim, cheng2024recy}.

Our work also aims to reduce material waste from PCBs. However, instead of focusing on new materials that may not be readily available to many, we seek to improve the workflow of the existing FR-4 substrate-based PCB manufacturing process. Our approach relies solely on off-the-shelf conductive epoxy and CNC engraving machines, which have become more affordable and widely available in makerspaces. As a result, our method has the potential to be adopted at scale.

\subsection{PCB Substrate Repair and Renewal}
Although PCBs are generally considered irreversible, several solutions have been proposed to repair minor errors or shorts. 
For example, jumper wires can restore electrical continuity between disconnected points~\cite{flyWire}, while conductive ink pens enable temporary, ad-hoc circuit repairs~\cite{trace_repair}. 
However, these methods are primarily effective for minor fixes, such as bridging gaps over short distances, and are not suitable for more complex repairs that require removing multiple conductors or altering component footprints and placements.

Several studies have investigated methods for fixing regional circuit errors. 
For example, Chen et al.~\cite{chen1991self} have developed a local electroplating technique to repair constrictions in copper traces.
Lim et al.~\cite{graphinePrint} have proposed repairing broken circuit traces using reduced graphene oxide on a laser direct writing platform.
Lange~\cite{lange2005pcb} has demonstrated the use of UV and IR lasers to trim fuzzy edges of conductor shapes on PCBs, reducing the defect rates in PCB products.
However, these approaches focus on repairing defects in PCB traces rather than addressing circuit design errors through rerouting or editing existing circuits.

Prior to our work, preliminary explorations have demonstrated the potential of using conductive filler deposition to modify or repair existing circuit diagrams on substrates.
For example, Self-healing UI ~\cite{self-healing} has introduced a composite material
capable of autonomously repairing circuit wiring made of multiwall carbon nanotubes by leveraging the dynamic cross-linking properties of polyborosiloxane polymers.
However, carbon nanotubes are hazardous and require specialized handling, and circuits made with this composite are limited to low-fidelity prototypes.
Circuit Eraser~\cite{circuiteraser} has proposed using a standard eraser to remove circuit traces printed with conductive ink, facilitating rapid iteration of circuit design.
Silver Tape~\cite{10.1145/3381013} enables circuit trace repair via tape transfer of inkjet-printed silver ink.
Furthermore, Marghescu et al. ~\cite{silverRepairCurrent} and Drumea et al. ~\cite{NiRepairCurrent} have evaluated the current-carrying capacity of sectional circuit traces made with nickel and silver paste, confirming the potential of PCB repair using conductive pastes.

Building upon previous research, we investigate the additive method of paste deposition as an alternative to the conventional subtractive PCB engraving process.
This approach enables the renewal of circuit boards originally fabricated using methods such as CNC engraving or photochemical etching. 
Furthermore, our method enables the editing of large conductive areas, allowing an existing PCB designed for a specific purpose to be repurposed for different projects. 
This, therefore, increases the opportunity to reuse otherwise wasted PCBs, reducing unnecessary e-waste.

\section{\pcbrenewal} \label{principle}
\pcbrenewal is a simple yet effective approach to repurposing PCB substrates that would otherwise be discarded. It helps reduce e-waste during PCB prototyping by enabling the correction of design errors, such as incorrect circuit trace connections or component misplacements, directly on faulty PCBs. Additionally, \pcbrenewal facilitates the reuse of obsolete outsourced PCBs, particularly open-source designs. By updating trace areas that are no longer needed, it provides retired PCBs with new functionalities.

\begin{figure}[h]
  
  \includegraphics[width=\columnwidth]{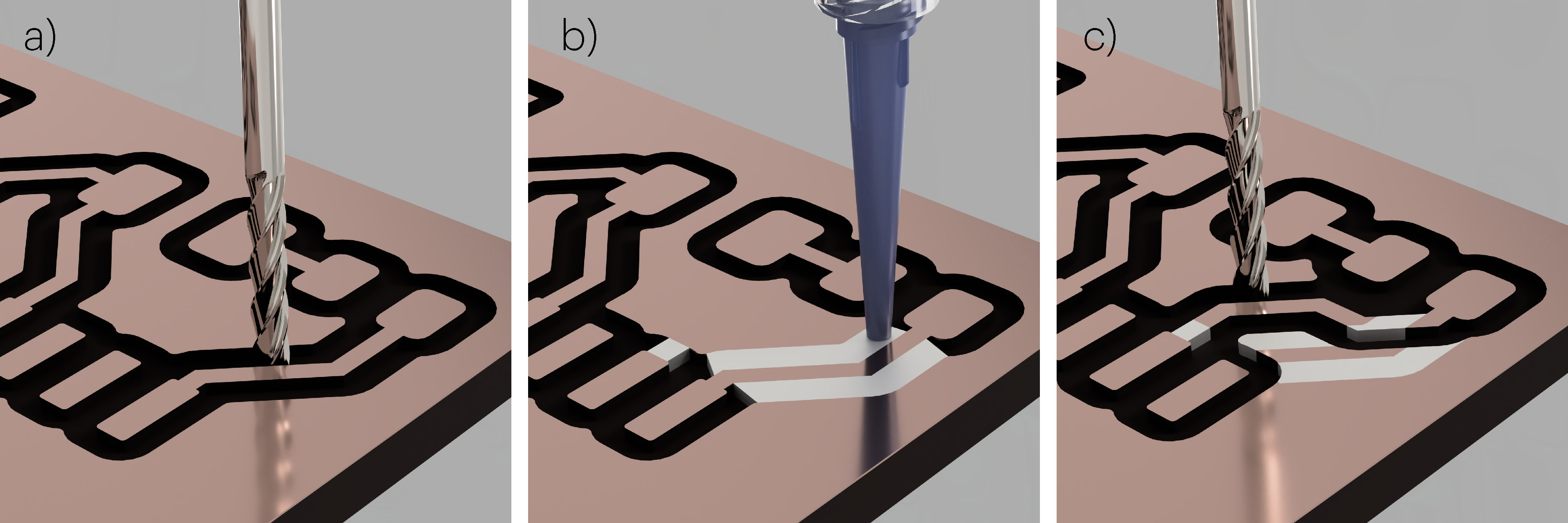}
  \caption{\pcbrenewal principle illustration: a) initial PCB engraved, b) selectively depositing conductive filler into isolation grooves, c) re-engraving new circuit trace.}
  \Description{}
  \label{fig:principle}
\end{figure}

As illustrated in Figure~\ref{fig:principle}, the core of the renewal process is the selective deposition of conductive filler material into isolation grooves to “erase” existing circuit traces or pads, allowing new conductive traces to be re-engraved. \pcbrenewal assumes access to conductive epoxy as a filler material and a CNC or laser-cutting machine for modifying the PCB substrate. To support this process, an EDA software plug-in (Section~\ref{software}) has been developed to compare new circuit designs with the existing layout and apply selective modifications only where necessary.

By preserving existing copper conductors and much of the fiberglass substrate, \pcbrenewal significantly reduces material waste, manufacturing costs, and energy consumption while maintaining a fabrication time comparable to producing a new PCB. Its core refilling and re-engraving processes are independent of board type, making it suitable for both in-house and factory-made PCBs, as well as single- and double-sided designs, though creating new vias for double-sided PCBs requires manual effort.

In the following sections, we use in-house PCBs with FR-4 substrates to explore key considerations and experiments related to \pcbrenewal. In Section~\ref{example}, we showcase examples of repairing and repurposing PCB substrates fabricated both in-house and through outsourced manufacturers.

\subsection{Material} \label{material}\label{fab}
The key to \pcbrenewal is refilling the isolation grooves of a PCB substrate to restore the conductive plane. 
This requires the conductive filler material to exhibit high conductivity, comparable to that of the original copper conductors. 
In addition, the filler material must form a robust bond with the PCB substrate while possessing physical properties that allow for controlled and precise deposition.

Our search for suitable materials began with solder wire and solder paste, widely accessible conductive materials known for their excellent electrical conductivity.
However, these materials are designed primarily to create strong metal-to-metal bonds between electronic components and copper circuit pads. 
Specifically, they exhibit high surface tension in their liquid state and are formulated to form metallurgical bonds exclusively with unoxidized metal surfaces~\cite{10.31399/asm.tb.ps.9781627083522}.
As fiberglass is inert to metallurgical bonding, solder tends to flow toward the copper surface rather than settling in isolation grooves.

In contrast to solder, conductive epoxy products are widely used in PCB screen printing and plotting processes. 
These polymer-based conductive epoxies exhibit high electrical conductivity for circuit traces and cure to a glassy state rather than transition to a high-surface-tension liquid, as is the case with solder. 
This property allows for uniform bonding to both metallic and inert substrates. 

Conductive epoxies are formulated with a variety of fillers, including silver, nickel, copper, carbon, and graphene. 
Notably, silver-based epoxies are available in single-part formulations that require no mixing and do not need specialized curing treatments, such as formic acid fumes, laser processing, or flash lamp exposure. 
Therefore, we surveyed a range of off-the-shelf, single-part, thermoset silver-based conductive epoxies, as shown in Table \ref{tab:epoxy}.

\begin{table*}[htb!]
  \caption{Silver Epoxies Surveyed}
  \label{tab:epoxy}
  \begin{tabular}{lcccc}
  \toprule
Name (code)         & Volume resistance (\SI{}{\micro\ohm\cdot\centi\meter}) & Viscosity (\SI{}{\pascal\cdot\second}) & Curing time (\SI{}{\minute}) & Curing temp (\SI{}{\celsius}) \\
    \midrule
Voltera Conductor 3 & 127                                          & \multicolumn{1}{l}{not reported}     & 15                                    & 170                                 \\
AA-duct 2979        & 30                                           & 65                                   & 15                                    & 150                                 \\
ACI FS0142          & 6                                            & 15                                   & 15                                    & 150                                 \\
DM-SIP-3072S        & 7.5                                          & 10                                 & 10                                    & 150                                 \\
Metalon® HPS-021LV  & 10.4                                         & 2.6                                  & 30                                    & 150  \\
\bottomrule
\end{tabular}
\end{table*}

\begin{figure}[h]
  
  \includegraphics[width=\columnwidth]{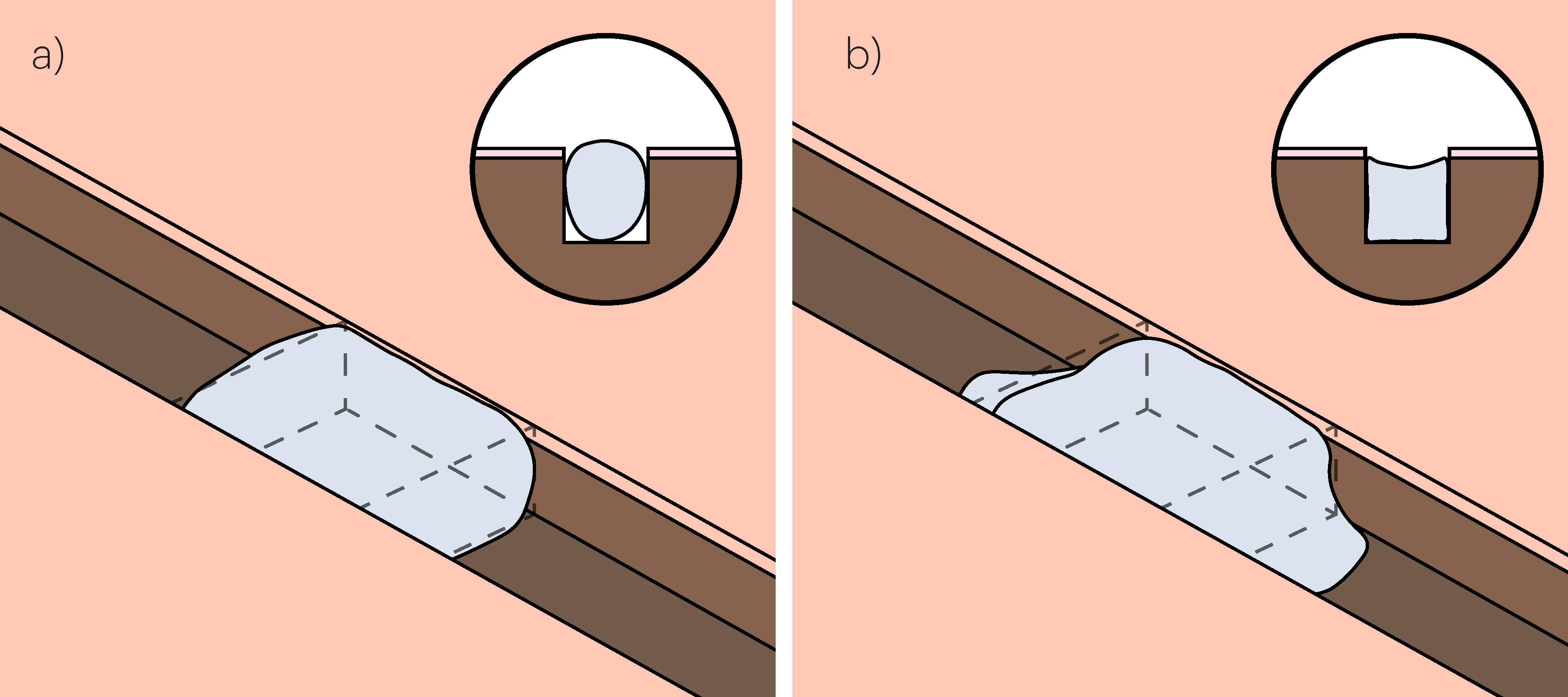}
  \caption{Illustration of conductive epoxy behavior at different viscosities: a) excessively high viscosity, b) excessively low viscosity.}
  \Description{}
  \label{fig:viscosity}
\end{figure}

We considered four technical criteria when selecting the appropriate conductive epoxy. 
First, the curing temperature must not exceed the maximum operating temperature of commonly used PCB substrates such as FR-4 (\SI{150}{\celsius} for Tg150 FR-4). 
Second, we prioritized materials with lower volume resistivity to maximize the current-carrying capacity of the traces passing through epoxy-filled areas. 
Therefore, we targeted a volume resistivity of the conductive epoxy that does not exceed \SI{10}{\micro\ohm\cdot\centi\meter}, which is within the same order of magnitude as copper.
Third, the viscosity of the material at room temperature is critical. Through empirical testing, we observed that excessive viscosity hinders efficient flow and proper filling of the filler material in tiny engraved grooves, resulting in poor mechanical bonding and unreliable electrical connections (Figure \ref{fig:viscosity}a).
On the other hand, excessively low viscosity causes the epoxy to flow away from the intended deposition areas or spread unevenly along the engraved grooves (Figure \ref{fig:viscosity}b).
Based on our experiments, we determined that a room-temperature viscosity of approximately 10-\SI{15}{\pascal\cdot\second} satisfies our requirements.
Fourth, to simplify the filler deposition process, we exclusively considered single-part conductive epoxy. This choice eliminates the need for mixing and minimizes material waste from residual mixtures.

These criteria led to the selection of two materials from the tested silver-based conductive epoxies: ACI FS0142 and DM-SIP-3072S. 
Based on material availability at the site the research was conducted, ACI FS0142 was chosen for all samples in this study unless otherwise noted. 
This heat-cured, single-part epoxy is specifically designed for PCB screen printing, has a viscosity of \SI{15}{\pascal\cdot\second} at room temperature, and cures at \SI{150}{\celsius} in 15 minutes.

Note that the goal of this search was to identify one conductive filler that meets our design requirements for \pcbrenewal. 
This survey is not exhaustive, and other materials may perform equally well or better.

\subsection{Fabrication Pipeline}
\begin{figure}[h]
  
  \includegraphics[width=\columnwidth]{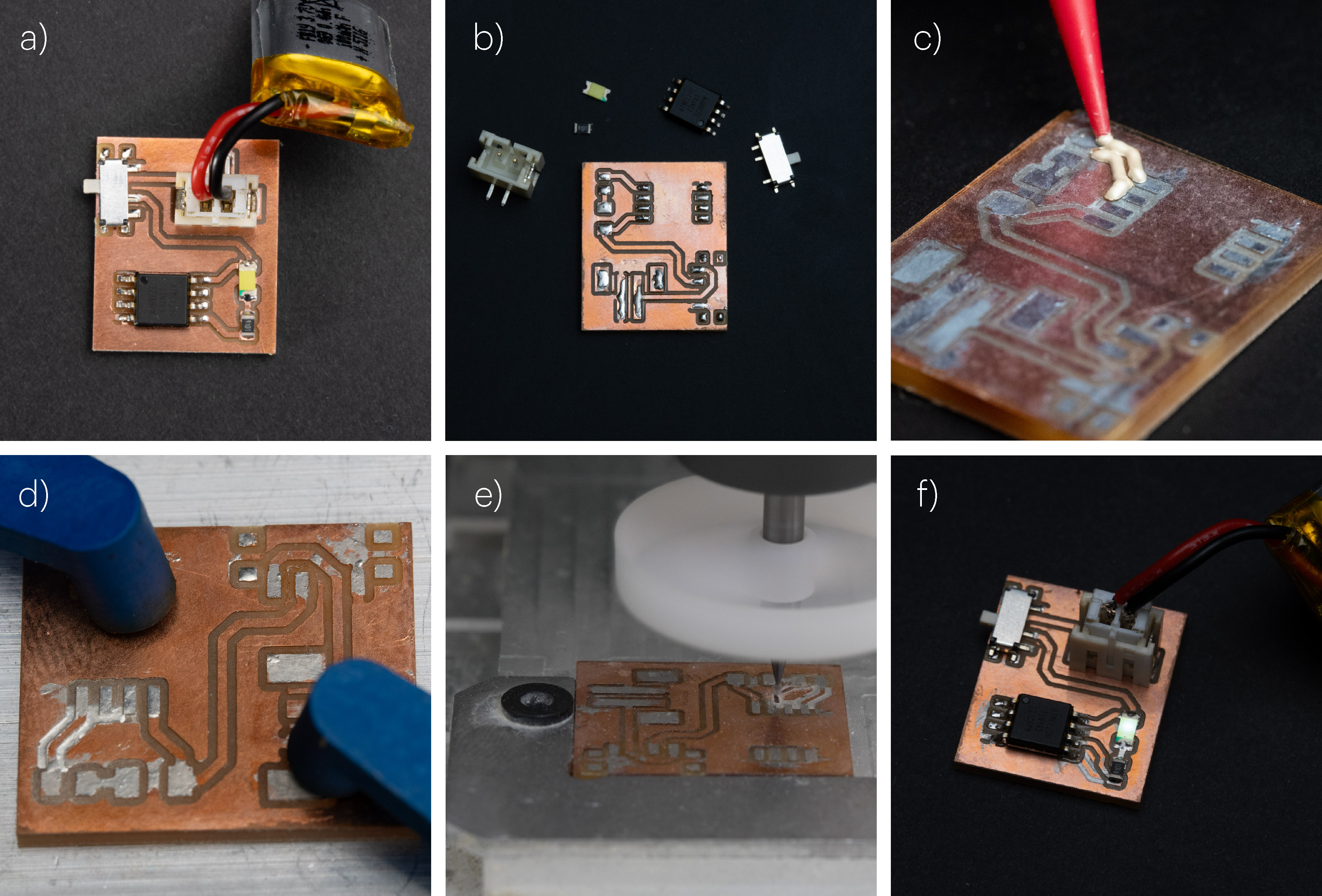}
  \caption{Fabrication workflow: a) old board, b) desoldering, c) manual epoxy deposition with a stencil,, d) epoxy curing, e) engraving new traces, f) new functional PCB with a modified trace.}
  \Description{}
  \label{fig:fab_step}
\end{figure}

The fabrication pipeline for renewing a PCB consists of four main steps: desoldering, depositing, curing, and engraving. 
We illustrate this process (Figure \ref{fig:fab_step}) by correcting an in-house PCB with a trace that was incorrectly connected due to a design error. 
Specifically, the example circuit includes an ATtiny85, a toggle switch, a JST connector, an LED, and a resistor that was mistakenly connected to the wrong pin of the ATtiny. 
During the renewal process, the incorrect trace is rerouted to connect to the correct pin, which is programmed to control the LED's blinking. 
As noted earlier, while we used a single-sided, CNC-milled PCB as the walkthrough example, our method is applicable to double-sided PCBs and those manufactured through outsourcing. %

\textit{Step 1 --- Desoldering}: \pcbrenewal begins with desoldering the components from the old PCB (Figure \ref{fig:fab_step}a-b). 
This step is essential because material deposition, curing, and new trace engraving can only be performed safely on a bare board. 
Although components far from affected areas might theoretically remain in place during small, localized modifications, we recommend fully removing all components. 
The heat curing process often reaches the solder's soaking temperature range, potentially compromising connection performance if components are left on the board.

\textit{Step 2 --- Depositing}: After desoldering, conductive epoxy is deposited into the engraved grooves to be restored. 
This process can be performed manually, similar to applying solder paste, or using a CNC machine with a paste extruder add-on. 
In our case, we use a syringe with a 23-gauge tapered blunt tip to manually deposit the conductive epoxy. 
An optional stencil can be generated from our software plugin (see Section~\ref{software}). 
The stencil profile features openings that align with the isolation areas to be restored (Figure \ref{fig:fab_step}c). 
When applying epoxy, it is important to ensure that there are no visible gaps between the epoxy and the adjacent copper to prevent open circuit spots on the updated board. 
Excess material can be removed manually before peeling the stencil off the board or with a CNC milling machine after the epoxy has cured.

\textit{Step 3 --- Curing}: Once the epoxy is applied, the board is cured by placing it in a convection oven or on a soldering hot plate.
We cure the epoxy at \SI{150}{\celsius} for 15 minutes using a hot plate (see Figure \ref{fig:fab_step}d). 

\textit{Step 4 --- Engraving}: After curing, the board is allowed to cool to room temperature before being placed on the CNC milling machine to engrave the updated traces (Figure \ref{fig:fab_step}e). 
An alignment bracket is used to position the bottom-left corner of the board at the machine origin. 
The engraving profile, obtained from the software plug-in, is then imported and aligned with the machine origin in the CNC control software. 
Since cured silver epoxy is softer than the FR-4 substrate, the engraving Gerber file and G-code can be generated using the same tooling and settings as a standard FR-4 PCB.
In this project, all samples are engraved using a Bantam Tools desktop CNC milling machine~\cite{Othermill}.

\section{Performance Characterization}
As \pcbrenewal introduces conductive materials other than copper for creating new PCB traces, it is essential to evaluate its electrical and mechanical performance to determine whether it can serve as a reliable iterative PCB making approach. 
To this end, this section outlines a series of experiments designed to evaluate \pcbrenewal’s performance, focusing on factors such as fabrication resolution in epoxy areas, electrical conductivity at copper-epoxy intersections, the current-carrying capacity of the traces, soldering performance, and the maximum number of renewal cycles achievable with this method.

\subsection{Fabrication Resolution}  \label{fabReso}

In CNC-engraved PCBs, the bonding strength between the conductive and dielectric layers is inversely related to the minimum trace width. 
Thinner traces are more prone to delamination from the fiberglass substrate. 
Consumer-grade CNC milling machines generally recommend trace widths of at least 10 mil (where 1 mil is one-thousandth of an inch or \SI{0.0254}{\milli\meter})~\cite{bantamTraces}. 
Renewed PCBs, which bond conductive epoxy to the fiberglass substrate through heat curing, may exhibit different bonding strengths compared to copper in standard FR-4. 
To determine the minimal engravable trace width for renewed PCBs, we conducted an experiment using varying trace widths in a conductive epoxy pour.

\begin{figure}[h]
  
  \includegraphics[width=\columnwidth]{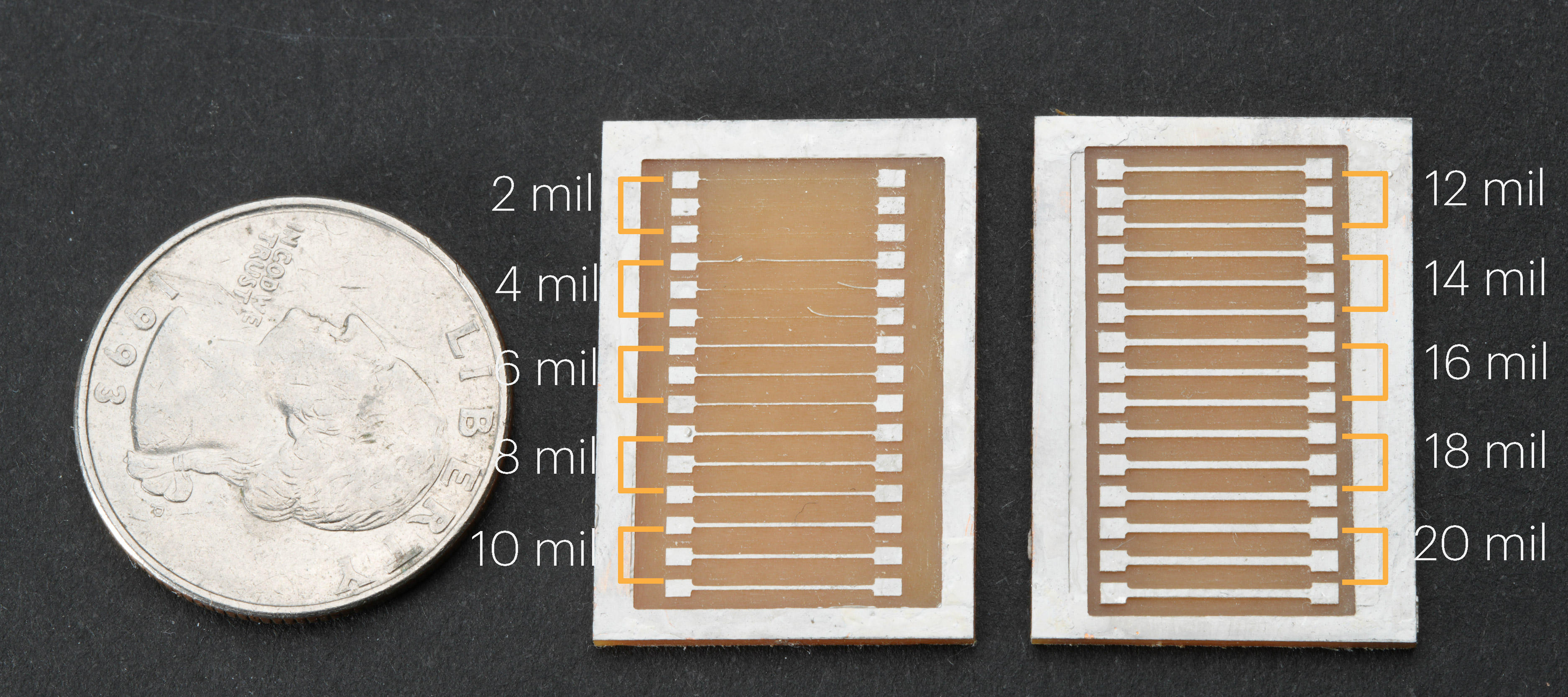}
  \caption{Fabrication resolution: trace engraving was attempted on a conductive epoxy pour at various trace widths.}
  \Description{}
  \label{fig:trace_width}
\end{figure}

We began by engraving a rectangular area on an FR-4 board to a depth of \SI{0.15}{\milli\meter}, which is the typical depth for creating PCBs with desktop CNC machines. 
The engraved area was then filled with conductive epoxy, leveled to flush with the surrounding copper, and cured on a hot plate.
Once the epoxy was fully cured, \SI{10}{\milli\meter} circuit traces with contact pads at both ends were engraved directly onto the epoxy surface.
The trace widths ranged from 2 to 20 mil, increasing in 2-mil increments.
Each width was tested three times, with the results shown in Figure~\ref{fig:trace_width}.
Traces narrower than 6 mil failed in all three attempts, while those 6 mil and above consistently succeeded, aligning with the recommended minimum trace width for CNC copper circuits.
In practice, we recommend designing circuit traces with the widest width that a design can accommodate to ensure optimal reliability.

\subsection{Electrical Conductivity}
A renewed PCB contains circuit traces made of silver epoxy or a hybrid of silver epoxy and copper.
To understand how variations in material composition affect trace conductivity, we conducted two sets of experiments.

\subsubsection{Conductive Epoxy Trace Conductivity} 

To evaluate the conductivity performance of the silver epoxy traces, we used traces with widths of 6 mil and above from the samples fabricated in Section~\ref{fabReso}.
Since the actual width of the engraved traces is influenced by manufacturing errors, we measured the actual trace width using a microscope stage, interpolating measurements to 0.1 mil.
The resistance of each trace was measured using a Keysight 3446SECU digital multimeter. 
The average measured trace width and resistance for each specified trace width were calculated from measurements taken across three individual traces.
The average trace widths were rounded to two decimal places, while the average resistance values were rounded to three decimal places, as presented in Table~\ref{tab:trace_cond}.

\begin{table}[h]
  \caption{Conductivity of Conductive Epoxy Traces}
  \label{tab:trace_cond}
  \begin{tabular}{ccccccccc}
    \toprule
    Nominal width (mil) & Measured width (mil) & Resistance (\SI{}{\ohm}) \\
    \midrule
    6 & 3.47 & 0.287 \\
    8 & 6.83 & 0.134 \\
    10 & 9.50 & 0.136 \\
    12 & 10.93 & 0.108 \\
    14 & 12.20 & 0.105 \\
    16 & 14.37 & 0.102 \\
    18 & 16.33 & 0.101 \\
    20 & 18.03 & 0.088 \\
  \bottomrule
\end{tabular}
\end{table}

As shown in the table, all the traces exhibit a resistance of less than \SI{0.3}{\ohm} per centimeter, with the majority below \SI{0.15}{\ohm}, making them suitable for implementing most low-frequency DC circuit functionalities.

\subsubsection{Material Interface Conductivity} \label{interface}

\pcbrenewal creates bonding seams between copper and epoxy, through which current flows. To assess the reliability of these seams, we conducted an experiment simulating real-world conditions to evaluate the quality of the connections at these points.

\begin{figure}[h]
  
  \includegraphics[width=\columnwidth]{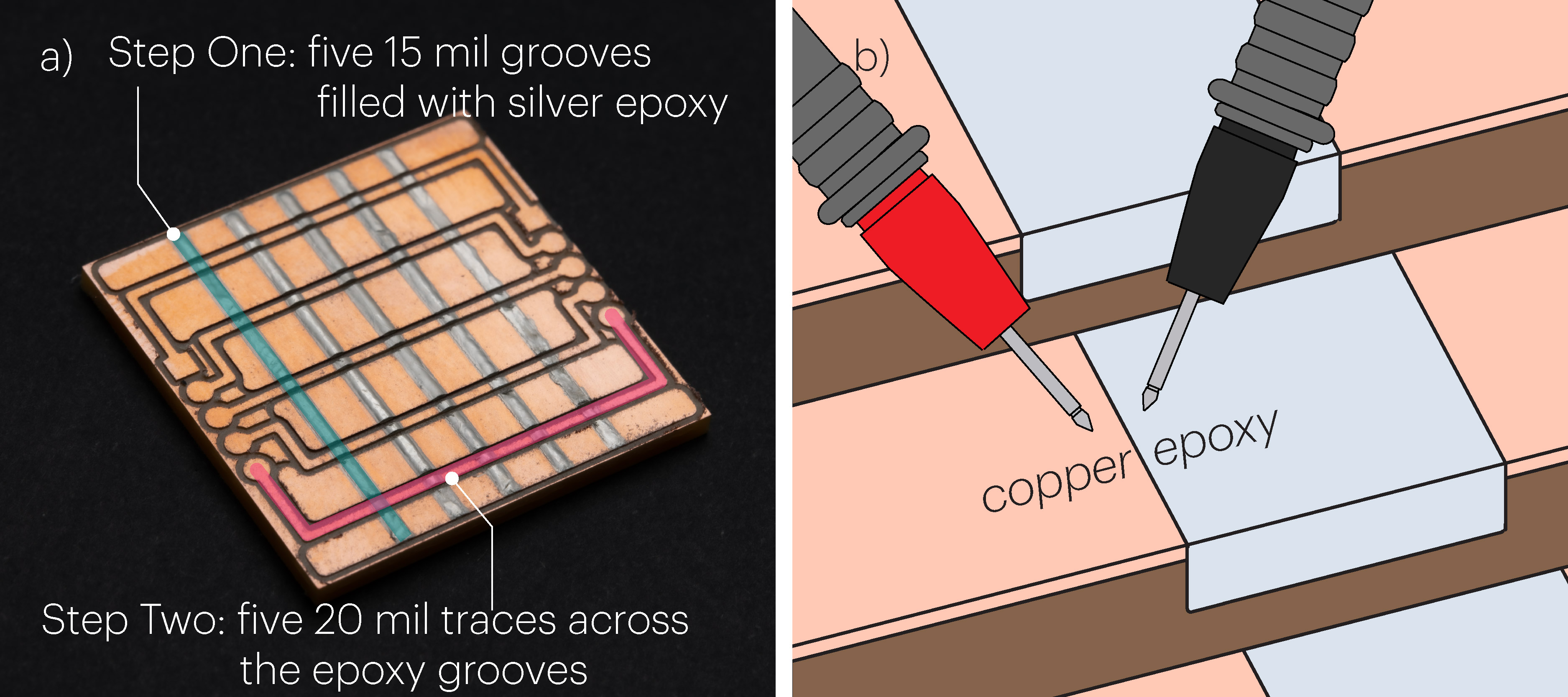}
  \caption{Material interface experiment: a) hybrid material traces (20 mil wide), b) illustration of measurement points.}
  \Description{}
  \label{fig:interface}
\end{figure}

We began with five parallel grooves, each 15 mil wide—the smallest typical square end mill diameter used for circuit boards with desktop CNC milling machines. 
Then, conductive epoxy was deposited in each groove. 
After curing, we engraved five 20 mil traces perpendicular to the grooves.
As a result, each trace contained 10 epoxy-copper bonding seams for investigation (Figure~\ref{fig:interface}a).
We measured the resistance of all 50 seams using a Keysight 3446SECU digital multimeter, probing as closely as possible to both sides of each seam (Figure~\ref{fig:interface}). 
The seams consistently exhibited a resistance of \SI{0.146}{\ohm} with a standard deviation of \SI{0.0345}{\ohm}, demonstrating that hybrid-material circuit traces can achieve electrical performance comparable to pure copper traces.

\subsubsection{Current Capacity}

Introducing an additional material into circuit trace formation can result in localized thermal accumulation due to uneven resistance. 
To evaluate the performance of conductive epoxy traces under high-current conditions, we tested the current-carrying capacity of the traces fabricated in Sections \ref{fabReso} and \ref{interface}. 
Fixed currents of \SI{1}{\ampere}, \SI{3}{\ampere}, and \SI{5}{\ampere} were applied to each trace using a bench power supply, and the temperature was monitored with a thermal camera. 
All experiments were conducted at a room temperature of \SI{22}{\celsius}.

\begin{figure}[h]
  
  \includegraphics[width=\columnwidth]{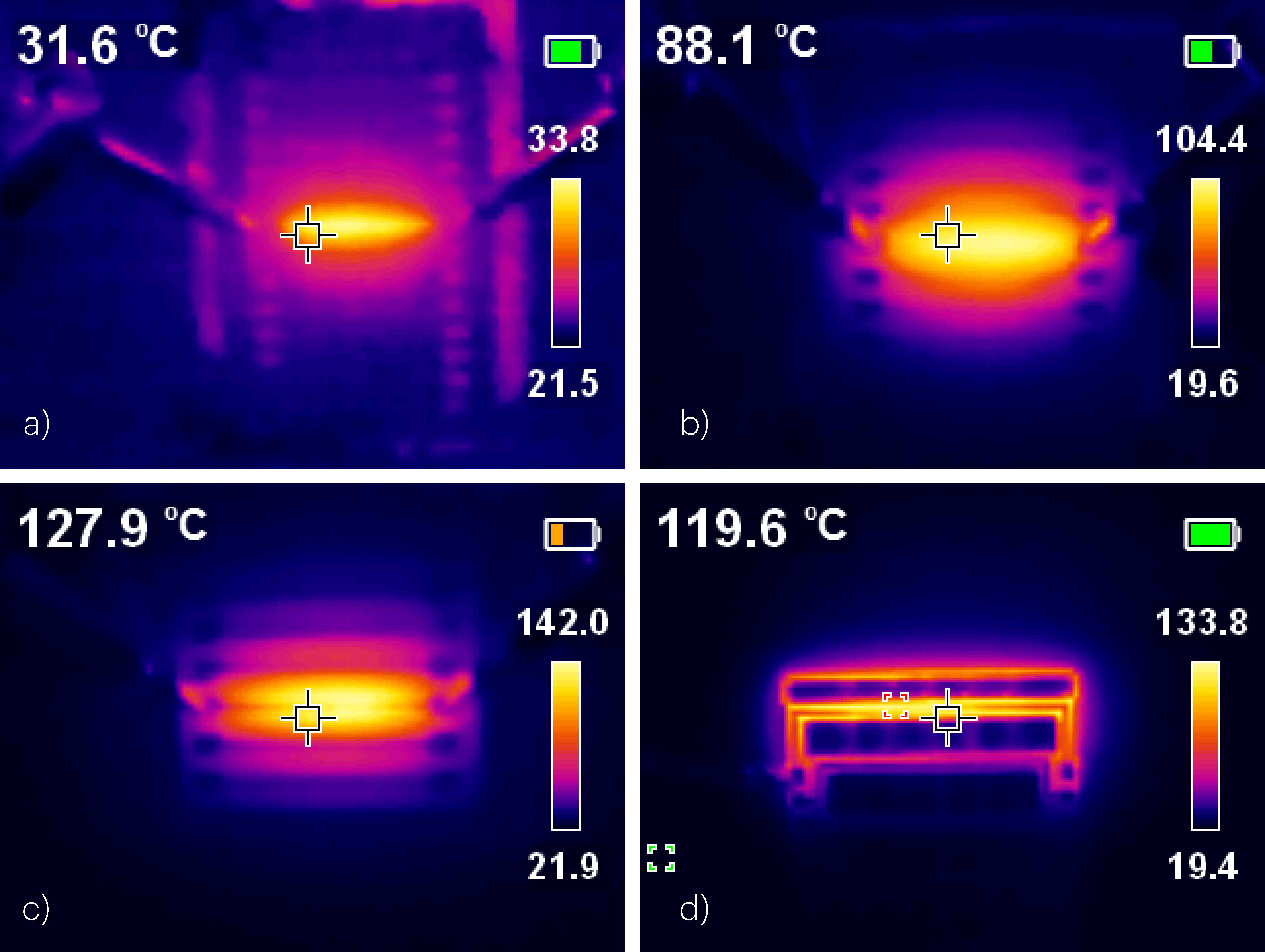}
  \caption{Current capacity experiment–thermal camera images of: a) 6 mil trace under 1A, b) 8 mil trace under 3A, c) 20 mil trace under 5A, d) hybrid material 20 mil trace under 5A.}
  \Description{}
  \label{fig:current_cap}
\end{figure}

We observed that the temperature increase of all traces remained below \SI{23}{\celsius} under a current of \SI{1}{\ampere} (Figure~\ref{fig:current_cap}a).
When subjected to \SI{3}{\ampere}, 6-mil traces fused within three seconds, while the remaining traces exhibited a maximum temperature rise of \SI{66}{\celsius} (Figure~\ref{fig:current_cap}b).
Under a \SI{5}{\ampere} load, traces narrower than 20 mils fused in five seconds.
However, the 20-mil traces remained functional, with a temperature increase below \SI{120}{\celsius}, which is within the \SI{150}{\celsius} Tg rating of the FR-4 board (Figure \ref{fig:current_cap}c).
These results indicate that traces produced by our method have sufficient current-carrying capacity for low-current DC signal circuits. 
For applications requiring currents above \SI{3}{\ampere}, a minimum trace width of 20 mils is recommended.

Furthermore, we observed that the hybrid traces fabricated in Section~\ref{interface} exhibited higher temperature increases at the conductive epoxy segments. 
However, at the same current levels, the temperature rise did not exceed that of traces made entirely from conductive epoxy (Figure~\ref{fig:current_cap}d).

\subsection{Solder Joint Performance}

\begin{figure}[h]
  
  \includegraphics[width=\columnwidth]{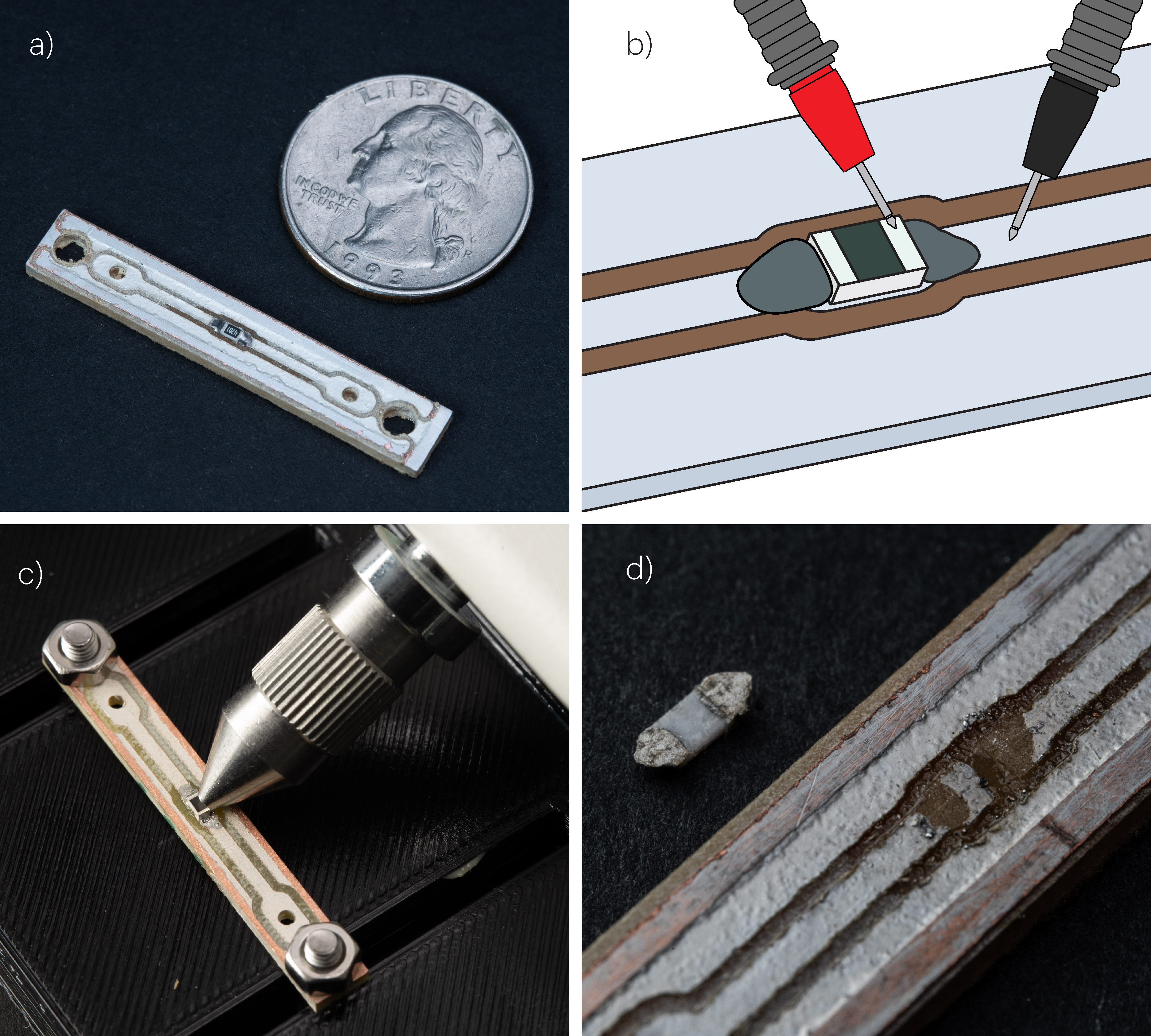}
  \caption{Solder joint experiment: a) an example of a solder joint experiment sample, b) the probing location adopted when measuring solder joint resistance, c) force gauge pressing on the soldered resistor at 30-degree angle, d) epoxy trace failure while the solder joint stays intact.}
  \Description{}
  \label{fig:soldering}
\end{figure}

In addition to circuit traces, PCB assemblies must ensure both conductivity and mechanical durability at solder joints.
The renewed PCB design will likely incorporate solder pads partially or entirely made of silver epoxy.
We investigated the conductivity and strength of the solder joint using 0805 resistors and their corresponding solder pads.
Following a process similar to that in Section \ref{fabReso}, we fabricated silver epoxy-based traces with solder pads designed for 0805 resistors.
The resistors were soldered (Figure \ref{fig:soldering}a) to six samples using low-temperature solder paste \cite{MG4902P}, as recommended by the silver epoxy manufacturer. 
Three samples were soldered using a hot plate, while the other three were soldered with a hot air blower.

The resistance of the solder joint was measured by probing the solder pad and the corresponding resistor terminal, using a 3446SECU digital multimeter (Figure \ref{fig:soldering}b). 
For comparison, we fabricated another set of samples on copper substrates with identical trace and pad geometry, soldering 0805 resistors using the same solder paste and soldering methods.
For both copper and epoxy pads, the hot air blower and hot plate methods produced similar solder joint resistance (Table \ref{tab:solderData}, rows 1 and 2).
The difference in resistance between solder joints on copper and epoxy pads was less than \SI{0.1}{\ohm}, a negligible value that does not affect the functionality of DC or AC signal circuits. 

\begin{table}
  \caption{Solder Joint Conductivity and Strength}
  \label{tab:solderData}
  \begin{tabular}{lccc}
    \toprule 
    Solder equipment  & Hot plate & Hot air & All samples \\ 
    \midrule
    Copper conductivity (\SI{}{\ohm})          & 0.17                      & 0.18                    & 0.18                            \\
    Epoxy conductivity (\SI{}{\ohm})           & 0.22                      & 0.27                    & 0.24                            \\
    Copper strength (\SI{}{\newton}) & 87.67                     & 71.40                   & 79.53   \\
    Epoxy strength (\SI{}{\newton}) & 39.32                     & 39.21                   & 39.27                           \\
  \bottomrule
\end{tabular}
\end{table}

In addition to conductivity, we used the same set of samples to evaluate the strength of the solder joints.
Pressure was applied to the soldered resistors at a 30-degree angle (Figure \ref{fig:soldering}c) until they detached from the solder pad. A DFS100 force gauge recorded the peak force value.
The samples soldered on epoxy pads broke off with an average force of \SI{39.27}{\newton}, with a standard deviation of \SI{16.16}{\newton}, while those soldered on copper pads withstood an average of \SI{79.53}{\newton}, with a standard deviation of \SI{17.51}{\newton}.
Note that for the silver epoxy samples, all break points occurred at the interface between the epoxy layer and the fiberglass, while the solder joints themselves remained intact (Figure \ref{fig:soldering}d). 
In practice, we recommend avoiding pure silver pads; however, if their use is necessary, increasing the pad size and the width of the connecting traces can help mitigate the risk of delamination.
Additionally, during testing, all connection points remained intact and functional, even after multiple drops from a height of \SI{1.5}{\meter}.

\subsection{Number of Renewal Iterations} \label{iteration}

In theory, an FR-4 substrate can be renewed indefinitely, provided that the newly engraved grooves consistently and completely remove the previous epoxy at the exact same Z-height.
However, in practice, achieving this level of machining precision is not feasible. 
To successfully renew a PCB, the engraving depth for new traces must be set deeper than the epoxy deposited in the previous iteration, which corresponds to the prior engraving depth.
Based on empirical results, we recommend that with each renewal iteration, the cutting depth be at least \SI{0.05}{\milli\meter} deeper than the previous one.

\begin{figure}[h]
  
  \includegraphics[width=\columnwidth]{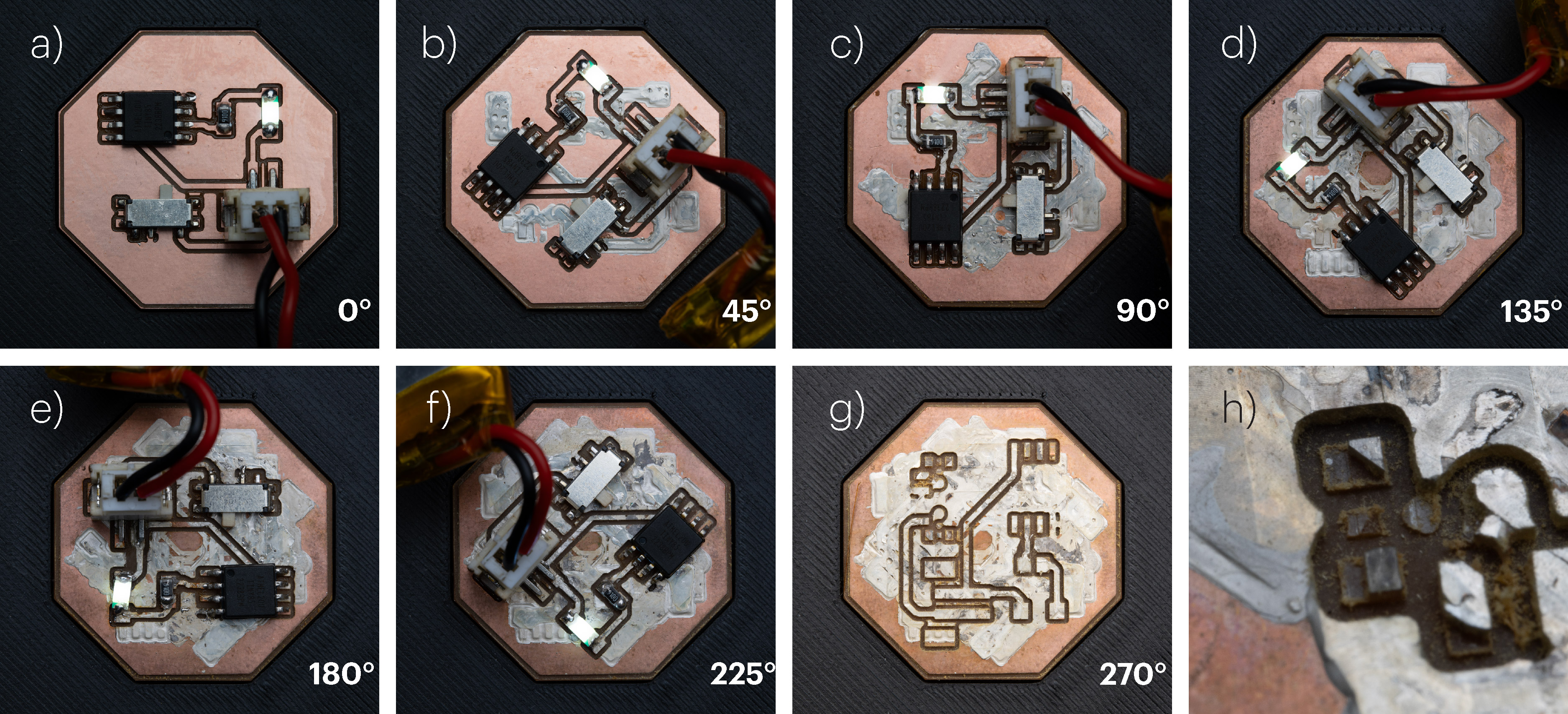}
  \caption{Multi-iteration renewal: a) original circuit, b)-f) second to sixth iteration of circuit modification, each rotated by 45 degrees counter-clockwise, g) the seventh iteration modification with broken traces and pads, h) zoom in view of a broken pad at the interface of copper and epoxy.}
  \Description{}
  \label{fig:iteration}
\end{figure}

As the cutting depth gradually increases with each renewal iteration on an FR-4 board, the trace is positioned progressively higher relative to the bottom of the isolation grooves, making the circuit traces more vulnerable during engraving.
We tested the maximum number of renewal iterations using an octagon-shaped PCB. 
The initial circuit consisted of an ATTiny85, a resistor, an LED, a JST connector for a LiPo battery, and a mini toggle switch. It was originally engraved with an isolation depth of \SI{0.15}{\milli\meter}. 
The minimum nominal trace width in the circuit was 16 mil (see Figure \ref{fig:iteration}a).
For each renewal iteration, we completely erased the old circuit by filling all engraved grooves with conductive epoxy, rotated the board by 45 degrees, and engraved the same circuit with an additional \SI{0.05}{\milli\meter} isolation depth (see Figure \ref{fig:iteration}b-f).
We found that the circuit traces remained functional until the seventh iteration, at which point small solder pads and traces began to break (Figure \ref{fig:iteration}g and h).

\section{Modeling the Sustainability Impact of \pcbrenewal} \label{sus_eval}

The primary goal of \pcbrenewal is to promote sustainable PCB making by enabling the reuse of PCB substrates. 
To fully understand its impact, a detailed evaluation is essential. Ideally, a lifecycle assessment (LCA)~\cite{hauschild2018life} would be conducted to comprehensively assess the environmental effects of \pcbrenewal. 
However, the variability of each renewal scenario makes it difficult to generalize its impact. For example, if a new circuit design shares no traces with the old one, the renewal process requires a near-complete removal of all old traces and the engraving of entirely new ones. Depending on the PCB size, this may result in a trade-off, where a minor reduction in FR-4 usage is offset by higher energy consumption for epoxy curing, potentially negating any environmental benefits when analyzed quantitatively.
Additionally, the lack of LCA data on most silver-based epoxy products further complicates a comprehensive LCA evaluation in practice.

To address this, we adopted the DeltaLCA framework~\cite{deltaLCA} and developed a quantitative comparison model that evaluates key sustainability metrics  commonly considered in LCA on a case-by-case basis. 
This model estimates and compares material usage, cost, time, and energy consumption between renewing a PCB and fabricating a new one from fresh FR-4. By analyzing these sustainability metrics, end-users can make informed decisions, determining whether renewing a PCB substrate is the more sustainable option or if fabricating a new one is preferable.

Note that while this section focuses on modeling the sustainability impact of \pcbrenewal, the model is also integrated into the software plug-in (Section~\ref{software}). 
As a result, all modeling parameters---such as deposition path length and trace contour length---can be directly extracted from PCB design profiles, enabling the automatic calculation of \pcbrenewal's sustainability impact for  each given PCB design.

\subsection{Modeling Material Usage and Cost Differences}
We chose to estimate material usage based on weight. 
While weight alone does not fully capture the material trade-offs between a PCB manufactured using the renewal approach and one made with new substrate, it provides the most practical basis for comparison, given the lack of comprehensive carbon footprint data for most silver-based conductive epoxies.
In \pcbrenewal, users are free to select any homemade or commercially available conductive filler following our guidelines. 
However, variations in the filler’s composition, manufacturing process, shipping distance, curing conditions, and cured material properties, along with other relevant factors, can significantly influence environmental impact metrics, including but not limited to carbon emissions, energy footprint, and toxicity.
For example, the energy footprint associated with mining and producing different metal elements used in conductive materials can vary by several orders of magnitude~\cite{TORRUBIA2023107281}.
Given these uncertainties, we provide material usage data in terms of weight as a reliable and conservative basis for further environmental impact modeling. 
This approach ensures consistency and prevents both overestimation and underestimation of the environmental implications of adopting \pcbrenewal.

For \pcbrenewal, the primary new materials introduced are conductive epoxy and, optionally, a deposition stencil sheet. 
The weight of epoxy required ($M_{e}$) can be estimated by multiplying the area of the isolation grooves to be filled ($A_{g}$) by the depth of the grooves from the previous engraving iteration ($d$) and the epoxy density ($\rho_{e}$). 
We offset the deposition depth by \SI{0.1}{\milli\meter} by default to account for excess material.
This parameter can be adjusted based on actual deposition needs.
The area of the stencil sheet ($A_{s}$) corresponds to the surface area of the previous board design ($A_{b\_old}$).

\begin{align*}
    M_{e} = \rho_{e} A_{g} d\\
    A_{s} = A_{b\_old}\\
\end{align*}

When calculating material usage for engraving a circuit on a new substrate, neither epoxy nor stencil material is involved. 
Instead, a fresh piece of FR-4 is used, with an area ($A_{FR-4}$) that matches the new board design ($A_{b\_new}$). 
\begin{equation*}
    A_{FR-4} = A_{b\_new}
\end{equation*}

We calculate the cost difference between the two methods (denoted as $P$) based on the unit prices ($p_u$) of each raw material and the estimated material usage. 
\begin{equation*}
P_{delta} = M_{e} p_{u\_e} + A_{s} p_{u\_s} - A_{FR-4} p_{u\_FR-4}
\end{equation*}
A negative value indicates monetary savings when using \pcbrenewal, while a positive value indicates additional costs.
Disposable hardware and equipment, such as tooling, double-sided tape, and glassware, are excluded from the material usage and cost estimation.

\subsection{Modeling Fabrication Time Differences}

The fabrication time for creating a circuit on a new substrate is the sum of trace engraving time, determined by the path length ($L_t$), and board outline cutting time.
The feed rate ($F_t$) depends on the engraving bit.
The number of passes is determined by the ceiling of the fraction of the engraving depth ($d_t$)---typically \SI{0.15}{mm}---and the stepdown ($\delta z_t$), which also depends on the engraving bit.
The board outline engraving time is calculated in the similar manner, based on the outline length ($L_o$), feed rate ($F_o$), board thickness as cutting depth ($d_o$), and stepdown ($\delta z_o$).
The total fabrication time can be estimated as follows:

\begin{equation*}
    T_{FR-4} = \frac{L_t}{F_t } \lceil{\frac{d_t}{\delta z_t}}\rceil+ \frac{L_o}{F_o }\lceil \frac{d_o}{\delta z_o} \rceil
\end{equation*}

The fabrication time in \pcbrenewal comprises several components: desoldering time, solder pad cleaning time, epoxy deposition time, epoxy curing time, engraving time, and an optional laser cutting time for manufacturing the deposition stencil. 
Desoldering time and solder pad cleaning time are highly dependent on the equipment used and the operator's skill level.
In practice, desoldering time ($T_{de}$) requires user estimation based on their specific scenario.
The default value for desoldering time is set to 1 minute, as all example circuits in our experiments were desoldered within this time frame using a hot plate.
The solder pad cleaning time ($T_{cl}$) is calculated as the number of solder pads on the old board ($n_{p}$) multiplied by the estimated cleaning time per pad ($t_{p}$).
Based on empirical experiments, the typical cleaning time per solder pad using a soldering iron is approximately 6 seconds. 
This value is set as the default, but users can adjust it to match their skill level.
Epoxy is deposited along the contours of the conductors designated for removal, with the extruder moving at a constant rate during deposition. 
The estimated deposition time ($T_{d}$) is calculated based on the total deposition path length ($L_{d}$) and the feed rate ($F_d$), which is set at \SI{3}{mm/s} for manual deposition.
\begin{equation*}
    T_{d} = \frac{L_{d}}{F_d}
\end{equation*}
Epoxy curing time ($T_c$) is a fixed duration specified in the conductive epoxy's datasheet.
Engraving time consists of the same two components as engraving a new board: trace engraving time and board outline cutting time. 
These are calculated using the methods described above, with the corresponding path lengths denoted as $L_t'$ for trace engraving and $L_o'$ for board outline modification cutting.
The primary difference lies in the trace engraving depth.
In \pcbrenewal, the new conductors must be engraved \SI{0.05}{\milli\meter} deeper than previous iterations (see Section \ref{iteration}).
Since the current renewal is the $n^{th}$ iteration, the engraving depth is:
\begin{equation*}
    d_t' = d_t + 0.05(n-1)
\end{equation*}
The stencil cutting time is estimated based on the contour length of the conductors to be removed ($L_s$) and the feed rate of a CO\textsubscript{2} laser cutter ($F_l$).
Hence, the time difference between renewing and engraving a new PCB is:
\begin{equation*}
    T_{delta} = T_{de} + n_{p} t_{p} + \frac{L_{d}}{F_d} + T_c + \frac{L_s}{F_l} + \frac{L_t'}{F_t } \lceil{\frac{d_t'}{\delta z_t}}\rceil -\frac{L_t}{F_t } \lceil{\frac{d_t}{\delta z_t}}\rceil + \frac{L_o'-L_o }{F_o}\lceil{\frac{d_o}{ \delta z_o}}\rceil
\end{equation*}

\subsection{Modeling Energy Consumption Differences}

The primary energy consumption arises from the epoxy heat curing process, as well as the power drawn by machinery for epoxy deposition, engraving, and stencil fabrication.
Energy consumption for desoldering, pad cleaning, deposition, engraving and stencil cutting is calculated by multiplying the estimated time for each stage by its respective power consumption.
Thus, the difference in energy consumption can be expressed as:

\begin{equation*}
\begin{split}
    E_{delta} & = 
    T_{de} P_{de} + n_{p} t_{p} P_i + \frac{L_{d}}{F_d} P_d + T_c P_c + \frac{L_s}{F_l} P_l
    \\
    & + (\frac{L_t'}{F_t } \lceil{\frac{d_t'}{\delta z_t}}\rceil -\frac{L_t}{F_t } \lceil{\frac{d_t}{\delta z_t}}\rceil + \frac{L_o'-L_o }{F_o}\lceil{\frac{d_o}{ \delta z_o}}\rceil) P_e 
\end{split}
\end{equation*}

where, $P_i$ denotes the power required by the soldering iron.

\section{Software} \label{software}

The \pcbrenewal software (open-sourced on GitHub\footnote{Software plug-in: \url{https://github.com/zyyan20h/PCBRenewal.git}}) serves three main purposes: visualizing and enabling direct comparison of two circuit designs, generating stencil profiles for epoxy filling and milling profiles for selective trace renewal, and automatically estimating the material usage, cost, time, and energy consumption savings or trade-offs of a given design. The software is developed as a plug-in for the open-source EDA software KiCAD. The plug-in uses KiCAD's Python bindings\footnote{KiCAD Python Bindings: \url{https://dev-docs.kicad.org/en/apis-and-binding/pcbnew/}} to access PCB data, shapely\footnote{shapely: \url{https://shapely.readthedocs.io/en/stable/}} for geometric operations, and wxPython\footnote{wxPython: \url{https://wxpython.org/index.html}} for the user interface.

\subsection{Software Plug-in Features}

The user interface includes a sequence of essential features: loading EDA files, aligning design layouts, selecting PCB layers for comparison, executing the comparison process, conducting sustainability analyses, and exporting cutting profiles. 
A responsive visualization panel remains active throughout the workflow, providing real-time updates based on user interactions to ensure immediate feedback.

\begin{figure}[h]
    \centering
    \includegraphics[width=\columnwidth]{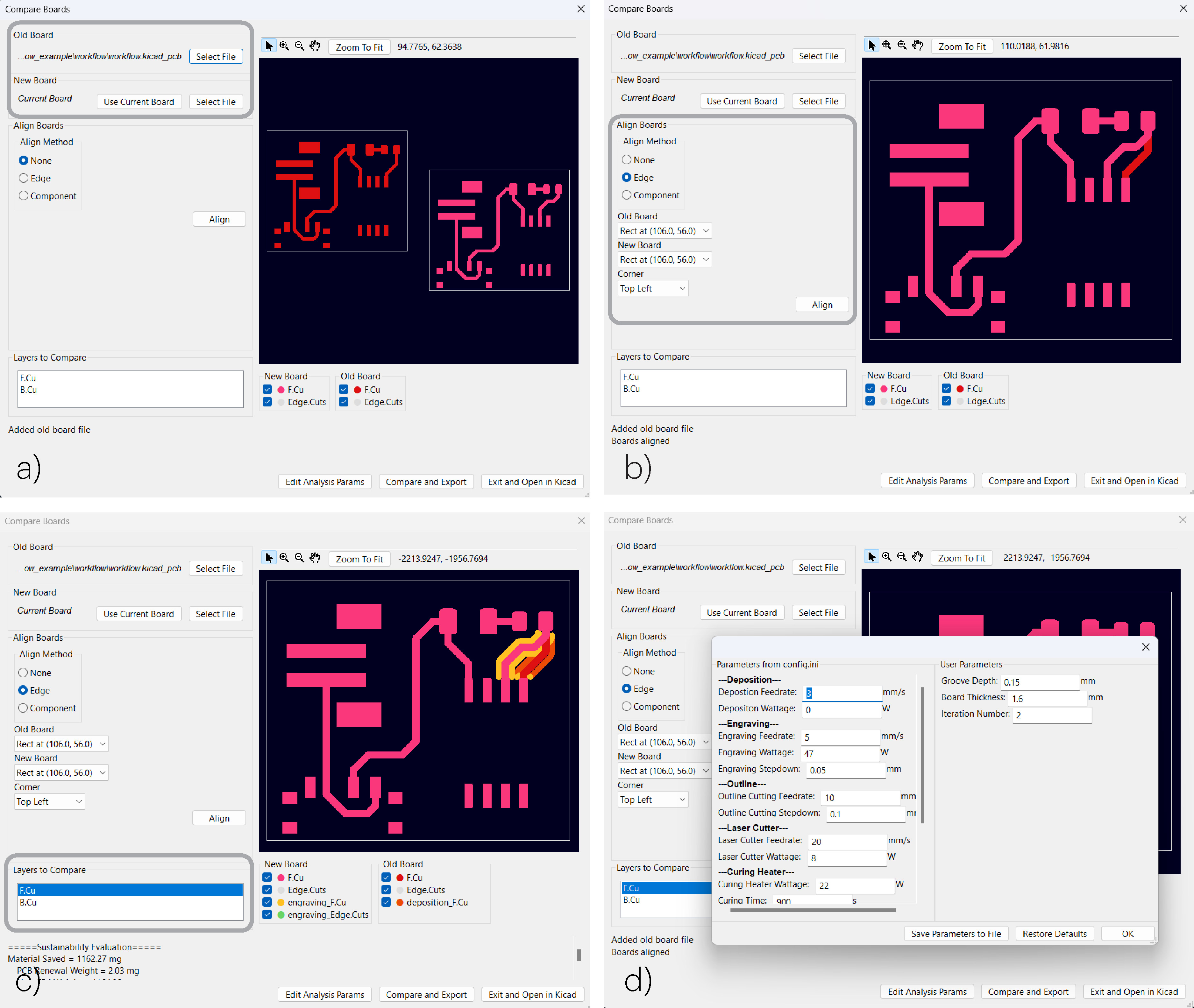}
    \caption{User Interface: a) importing boards, b) board alignment, c) board comparison, d) sustainability analysis parameters made customizable for different machine and tooling adoption.}
    \Description{}
    \label{fig:UI}
\end{figure}

\paragraph{Board Comparison.}
Our software allows users to load two KiCad PCB designs for comparison (Figure \ref{fig:UI}a).
Because the designs may vary in size and position, an optional feature enables users to align them using selected reference points, such as the corners of board outline bounding boxes or the geometric centers of electronic component footprints (Figure \ref{fig:UI}b).
Once aligned, the software executes the comparison algorithm in the background and displays the results in the visualization window. 

\paragraph{Output and Analysis}
After comparison, the software automatically exports the stencil profile and engraving pattern as fabrication inputs to the same directory as the ``new board'' design.
While exporting these files, the software also performs a sustainability analysis for the given renewal scenario and displays the result in the log at the bottom of the plug-in interface.
Based on these results, users can decide whether to proceed with \pcbrenewal or create a new PCB from scratch.
Calculation parameters are initialized with default values that match the machines and tools used in our demonstrations.
Users can reconfigure these parameters in a pop-up window by clicking the ``Edit Analysis Params'' button (Figure \ref{fig:UI}d).

\subsection{PCB Design Comparison Algorithms}

The circuit design comparison results are used as both fabrication input and sustainability impact analysis data.
This process requires highly accurate output to ensure minimal fabrication errors and reliable analysis results.
To achieve a precise comparison between two KiCad board designs, we developed a custom algorithm that extracts board information from KiCad and converts it into vector-based geometries.

We used KiCAD's Python bindings to access the board information.
Every PCB component (e.g. pads, tracks, holes) incorporated in our comparison algorithm is represented as a user-defined instance to preserve the integrity of the original data.
Each board is represented by an instance of a custom \verb|Board| class. A \verb|Board| instance contains a collection of nets---groups of electrical nodes (or pins) and tracks that are electrically connected on copper layers. 
These nets are stored in a nested hash map, $H$, where each key corresponds to a layer names (e.g., \verb|F.Cu| for the front layer or \verb|B.Cu| for the back layer), and each key points to a list of nets present on that layer.
Nets are represented by a custom class, and each net instance contains a list of tracks, a list of pads, and the layer name to which it belongs. 
When the boards are imported, we initialize the board instances according to the layer and net information retrieved from KiCad.
We refer to the two board instances as $B_{old}$ and $B_{new}$, and their respective net hash maps as $H_{old}$ and $H_{new}$.

The comparison is carried out in two steps. 
First, all nets from the old board are compared against each net of the new board. 
This step identifies nets with identical geometry and position, which remain unchanged and can be excluded from further comparison.

Next, the remaining nets in both board instances are converted into flat polygons and subjected to Boolean union operations within each board.
A second round of comparison is then performed on the resulting compound polygon outlines, producing the final comparison results.

\paragraph{Net Wise Comparison.}
This step takes $H_{old}$ and $H_{new}$, and generates two new hash maps, $H_{old\_unique}$ and $H_{new\_unique}$, each containing nets with unique geometries from their respective boards.\footnote{We use the notation hashmap[key] $\leftarrow$ value to represent inserting or updating a value associated with a specific key in the hash map, mirroring Python's dictionary syntax.}
If the two boards share no common nets, then $H_{old\_unique}$ will contain all the net instances from $B_{old}$ and $H_{new\_unique}$ will contain all the conductors from $B_{new}$.
The pseudocode block below illustrates the pairwise comparison of each list of nets within the corresponding layers of the boards.

\RestyleAlgo{ruled}

\begin{algorithm}
\caption{\textbf{compareNets($H_{old}$, $H_{new}$, $S$)}}\label{alg:com_conductors}

\textbf{in:} Hash map of nets on the old board $H_{old}$, Hash map of nets on the new board $H_{new}$, Layers selected for comparison $S$

\textbf{out:} Hash map of unique old nets $H_{old\_unique}$, Hash map of unique new nets $H_{new\_unique}$

\textbf{local:} Flag denoting whether a net in the old board has an identical match in the new board $F$

\begin{enumerate}
    \item $H_{old\_unique} \leftarrow$ empty hash map 
    \item $H_{new\_unique} \leftarrow$ empty hash map 
    \item \textbf{for each} layer $L$ in $S$ \textbf{do}: 
    \item \quad \quad $H_{old\_unique}[L] \leftarrow$ empty list
    \item \quad \quad $H_{new\_unique}[L] \leftarrow$ $H_{new}[L]$
    \item \quad \quad \textbf{for each} old net $N_{old}$ in $H_{old}[L]$ \textbf{do}:
    \item \quad \quad \quad \quad $F \leftarrow$ FALSE
    \item \quad \quad \quad \quad \textbf{for each} new net $N_{new}$ in $H_{new\_unique}[L]$ \textbf{do}:
    \item \quad \quad \quad \quad \quad \quad \textbf{if} $N_{old} = N_{new}$ \textbf{then:}
    \item \quad \quad \quad \quad \quad \quad \quad $F \leftarrow$ TRUE
    \item \quad \quad \quad \quad \quad \quad \quad \textbf{Pop} $N_{new}$ from $H_{new\_unique}[L]$
    \item \quad \quad \quad \quad \quad \quad \quad \textbf{Exit loop}
    \item \quad \quad \quad \quad \textbf{if not} $F$ \textbf{then}:
    \item \quad \quad \quad \quad \quad \quad \textbf{Append} $N_{old}$ to $H_{old\_unique}[L]$
    \item \textbf{Return} $H_{old\_unique}$, $H_{new\_unique}$
\end{enumerate}

\end{algorithm}

When comparing two nets (line 9 in Algorithm \ref{alg:com_conductors}), we verify that the position and geometry of all pads and tracks in both nets are identical.

\paragraph{Geometric Comparison}

In this step, we convert all remaining unique nets into polygons for further comparison.
Algorithm \ref{alg:create_paths} details the parsing process for these remaining nets within a single board.

\RestyleAlgo{ruled}

\begin{algorithm}

\caption{\textbf{createPaths($H$, $S$)}}\label{alg:create_paths}

\textbf{in:} Hash map of nets $H$, Layers to compare

\textbf{out:} Hash map of paths $P$

\textbf{local:} offset outline of an individual net $p_{net}$, compound geometry of all net in a layer $p$

\begin{enumerate}
    \item $P \leftarrow$ empty hash map 
    \item \textbf{for each} layer $L$ in $S$ \textbf{do}:
    \item \quad \quad $p \leftarrow$ blank shape \\ \tcp{place holder for the Boolean union paths}
    \item \quad \quad \textbf{for each} net $N$ in $H[L]$ \textbf{do}:
    \item \quad \quad \quad \quad $p_{net} \leftarrow$ offset outline of $N$
    \item \quad \quad \quad \quad $p \leftarrow$ \textbf{Boolean union} of $p$ and $p_{net}$
    \item \quad \quad $P[L] \leftarrow p$
    \item \textbf{Return} $P$
\end{enumerate}

\end{algorithm}

For each net---whether it is to be removed from the old board design or engraved into the new one---the fabrication process focuses on the isolation area outside that net, either covering it or removing materials.
The minimum width of the isolation area is usually defined in the design rule checking (DRC) configuration.
To determine the midline of the isolation area, we offset the outlines of each net by half of the minimum isolation width.
This midline conservatively represents any possible machining or deposition path outside the net.
Within each board layer, we then compute the Boolean union of all polygons generated for the leftover nets in that layer, and store the resulting path in a new hash map $P$.

Using Algorithm \ref{alg:create_paths}, we parse the leftover nets across all layers in both boards. We then apply Boolean subtraction between the parsing results of each layer from each board (Algorithm \ref{alg:comp_paths}).
This process yields paths for deposition ($D_{path}$) and engraving ($E_{path}$), each having a equal to the minimum isolation width defined in DRC.

\begin{algorithm}
\caption{\textbf{comparePaths($H_{old}$, $H_{new}$, $S$)}}\label{alg:comp_paths}

\textbf{in}: Hash map of nets on the old board $H_{old}$, Hash map of nets on the new board $H_{new}$, Layers selected for comparison $S$

\textbf{out:} Hash map of paths to deposit $D_{path}$, Hash map of paths to engrave $E_{path}$

\begin{enumerate}
    \item $P_{old} \leftarrow$ createPaths($H_{old}$, $S$)
    \item $P_{new} \leftarrow$ createPaths($H_{new}$, $S$)
    \item $D_{path} \leftarrow$ empty hash map 
    \item $E_{path} \leftarrow$ empty hash map
    \item \textbf{for each} layer $L$ in $S$ \textbf{do}:
    \item \quad \quad $D_{path}[L] \leftarrow$ \textbf{Boolean subtraction} of $P_{old}[L]$ and $P_{new}[L]$
    \item \quad \quad $E_{path}[L] \leftarrow$ \textbf{Boolean subtraction} of $P_{new}[L]$ 
    and \\ $P_{old}[L]$
    \item \textbf{Return} $D_{path}$, $E_{path}$
\end{enumerate}
\end{algorithm}

Note that, between individual renewal iterations, certain traces and pads from the old board do not need to be ``erased'' if the corresponding area is not utilized in the new board design.
However, it is uncertain whether future iterations will make use of the areas these traces and pads occupy.
To preserve the potential for all future renewal iterations, our software, by default, ``erases'' all undesired nets from the old board.

\paragraph{Support of Vias and Through-Hole Components.}

Vias are compared within their own category across the two boards.
When a via from the old board is no longer used in the new design, it is replaced with a hole in the engraving profile.
These holes, along with existing holes in the old board designed for through-hole components and mechanical assembly, are considered outside board usable profile and do not support new traces and pads.
If any new traces or pads overlap with these areas, our software will generate an error message and a corresponding visualization layer in yellow color to alert the user.

\paragraph{Outline Comparison.}

In addition to comparing the copper layers, the plugin also compares the board outlines. 
It does this by by converting the board outlines into polygons and performing a Boolean subtraction on those polygons.
This results in a polygon that serves as a guide for trimming the old board to convert it into the new one.

The plugin uses the shapely python library to perform geometric parsing and Boolean operations.

\section{Example \pcbrenewal Scenarios} \label{example}

In this section, we present a series of walkthrough examples. Sections~\ref{cam_roller} through ~\ref{espboy} showcase a single CNC-milled substrate being reused across four design iterations within three distinct projects. 
Section~\ref{outsourced_pcb} further demonstrates that \pcbrenewal can be applied to factory-made, double-layer PCBs.
These examples highlight how \pcbrenewal facilitates local alterations to circuit traces and board outlines, enabling error correction and functional updates. Additionally, they demonstrate the versatility and range of electrical functionalities achievable with these updated hybrid material circuits.

We report the sustainability analysis results for each example.
For trace engraving, we used a 1/64-inch square end mill, while a 1/32-inch square end mill was used for outline engraving. The corresponding tooling parameters were applied to estimate the fabrication time.
Additionally, we measured the average power consumption of our machines using an appliance wattage monitor. During operation, the CNC machine consumes approximately \SI{47}{\watt} for engraving, the hotplate averages \SI{22}{\watt} for desoldering, the solder iron used for pad cleaning consumes \SI{21.5}{\watt}, the laser cutter requires \SI{8}{\watt} for stencil cutting, and the heater operates at an average of \SI{22}{\watt} during the heat-curing process.
We set the desoldering time to 1 minute and the cleaning time for each solder pad to 3 seconds.
These values are used as inputs for energy estimation. 
The standardized analysis data are visualized in radar graphs for each renewal iteration.

\begin{figure*}[p]
  
  \includegraphics[width=0.75\textwidth]{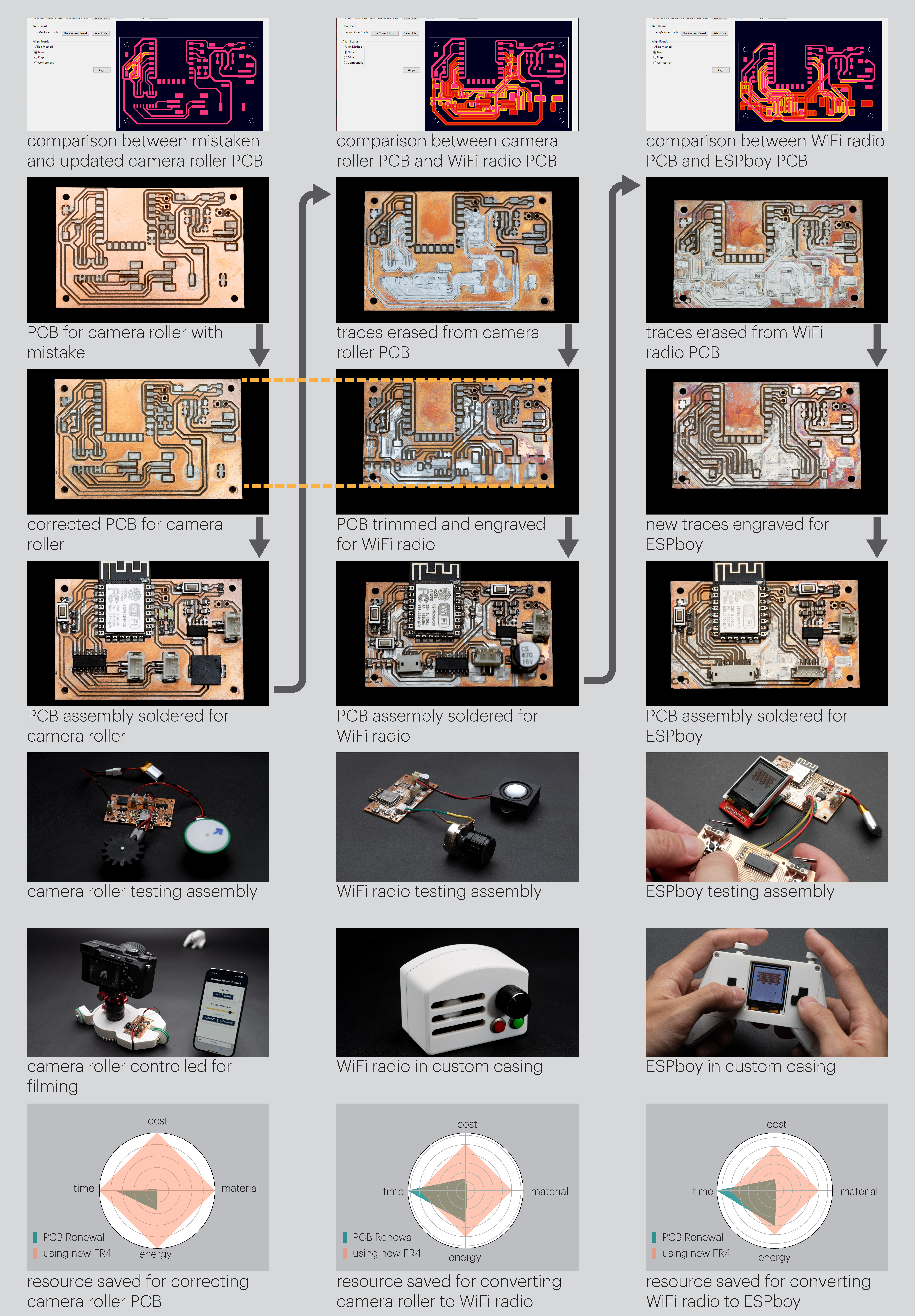}
  \caption{Multi-iteration renewal on a single piece of FR-4. Each column presents the software-generated PCB comparison result, the renewal process, the prototype assembly, and the amount of resources saved. Left column: correction of a mistaken connection in the camera roller PCB. Middle column: trimming the board size and modifying part of the camera roller circuit for the WiFi radio prototype. Right column: converting the WiFi radio circuit into the ESPboy motherboard, along with a daughterboard to expand functionality.}
  \Description{}
  \label{fig:examples}
\end{figure*}

\subsection{Iteration One and Two --- Camera Roller} \label{cam_roller}

In this example, we created a camera roller designed to achieve fluid, dynamic shots, such as tracking, panning, and dollying. 
The original circuit board was developed to control two DC gear motors using an ESP8266 microcontroller. 
However, an error was identified in the ESP8266 accessory circuit—its enable pin requires an external pull-up resistor when resetting the board or entering download mode, preventing us from uploading code to the ESP8266.

To correct this, we needed to add a resistor and connect it to two existing conductors, which also required relocating some components and traces. 
In a conventional PCB prototyping process, this would have required manufacturing an entirely new PCB, as traces cannot be easily altered or added.

With \pcbrenewal, however, we were able to make these minor adjustments directly on the existing prototype. 
This enabled us to implement the necessary modifications without the waste of materials or energy required to fabricate a new board. 
The corrected PCB now functions as intended, allowing control code to be uploaded and ensuring smooth operation of the camera roller (see the left column of Figure \ref{fig:examples}). 

Between the first and second design iterations, \pcbrenewal allowed us to save \SI{6402.90}{\milli\gram} of FR-4, \SI{71.91}{\kilo\joule} of energy and \SI{15.25}{\minute} in fabrication time, while consuming only \SI{4.06}{\milli\gram} of silver epoxy, reducing the cost of raw material by 98.4\%.

\subsection{Iteration Three --- WiFi Radio}

With the camera roller design finalized, the prototype PCB was no longer needed. 
However, much of its circuitry, especially the sections supporting the ESP8266 microcontroller, remained potentially useful for other projects. 
Instead of discarding the entire board, we selectively removed and updated only the necessary components of the camera roller PCB, repurposing it for a new project.

In this case, we transformed the otherwise obsolete camera roller PCB into a WiFi radio controller while retaining much of the original microcontroller circuitry. 
The modifications mainly involved swapping out the motor driver and connectors for an audio amplifier, speaker connections, and a potentiometer. 
We designed the new circuit layout using KiCAD and utilized the \pcbrenewal plug-in to evaluate the sustainable impact of updating the old board. 
We then physically implemented the updated PCB by selectively removing and updating the traces and pads, as well as reducing the board size to fit the new radio design. 
The renewal process is documented in the middle column of Figure \ref{fig:examples}. 

In addition to demonstrating how \pcbrenewal can support the prototyping of a complete new project using an obsolete PCB, this WiFi radio example also showed that the renewed PCB, with circuit traces made from hybrid materials, could support audible-frequency data transmission while maintaining low noise levels. 

In this design iteration, renewing the PCB allowed us to save \SI{5602.15}{\milli\gram} of FR-4 and \SI{32.03}{\kilo\joule} of energy while consuming only \SI{105}{\milli\gram} of silver epoxy, reducing material cost by 74.6\%.
The fabrication time is comparable to engraving a new piece of FR-4, with \pcbrenewal taking only 3.89 minutes longer despite the additional desoldering and cleaning steps.

\subsection{Iteration Four --- ESPBoy Game Console}\label{espboy}
One FR-4 board can undergo multiple iterations across different projects. Here, we demonstrate that the same PCB substrate can be reused for yet another new project, even after three prior iterations.

Specifically, we retrofitted the previous WiFi radio controller into a game console based on the open-source ESPBoy design~\cite{espboy}. 
In this iteration, we repurposed the WiFi radio circuit as the motherboard of the ESPBoy assembly, retaining the ESP8266 circuitry and adding two multi-pin JST connectors. 
Additionally, we fabricated a daughterboard that hosts an OLED display and joystick controls, serving as the console’s main input and output interface. These components were positioned ergonomically to ensure comfortable operation. The multi-pin JST connectors linked the ESP8266 motherboard with the daughterboard.

The updated motherboard effectively handled high-frequency, real-time data transmission, as demonstrated by the I²C communication at \SI{100}{\kilo\bit/\second} between the microcontroller, the display, and the GPIO extender that processed the button inputs.
The renewal process is documented in the right column of Figure \ref{fig:examples}.

While this iteration introduced an additional PCB, we still reduced material waste by largely reusing the original PCB as the motherboard of the ESPBoy game console. Specifically, we saved \SI{5608.24}{\milli\gram} of FR-4 and \SI{25.99}{\kilo\joule} of energy while consuming only \SI{98.91}{\milli\gram} of silver epoxy, reducing material costs by 87.5\%.
The fabrication time remains comparable to engraving a new FR-4 board, with a difference of less than \SI{5}{\minute}.

\subsection{Renewing an Outsourced PCB}\label{outsourced_pcb}
While previous examples showcased how \pcbrenewal reduces material waste for CNC-milled PCBs, its versatility extends to factory-made PCBs, such as those ordered online or found in commercial electronic devices. In this example, we repurposed a digital LED watch PCB, manufactured as a double-layer board with a solder mask by a small-batch PCB producer, into a PCB for an interactive cat toy.

\begin{figure}[h]
  
  \includegraphics[width=\columnwidth]{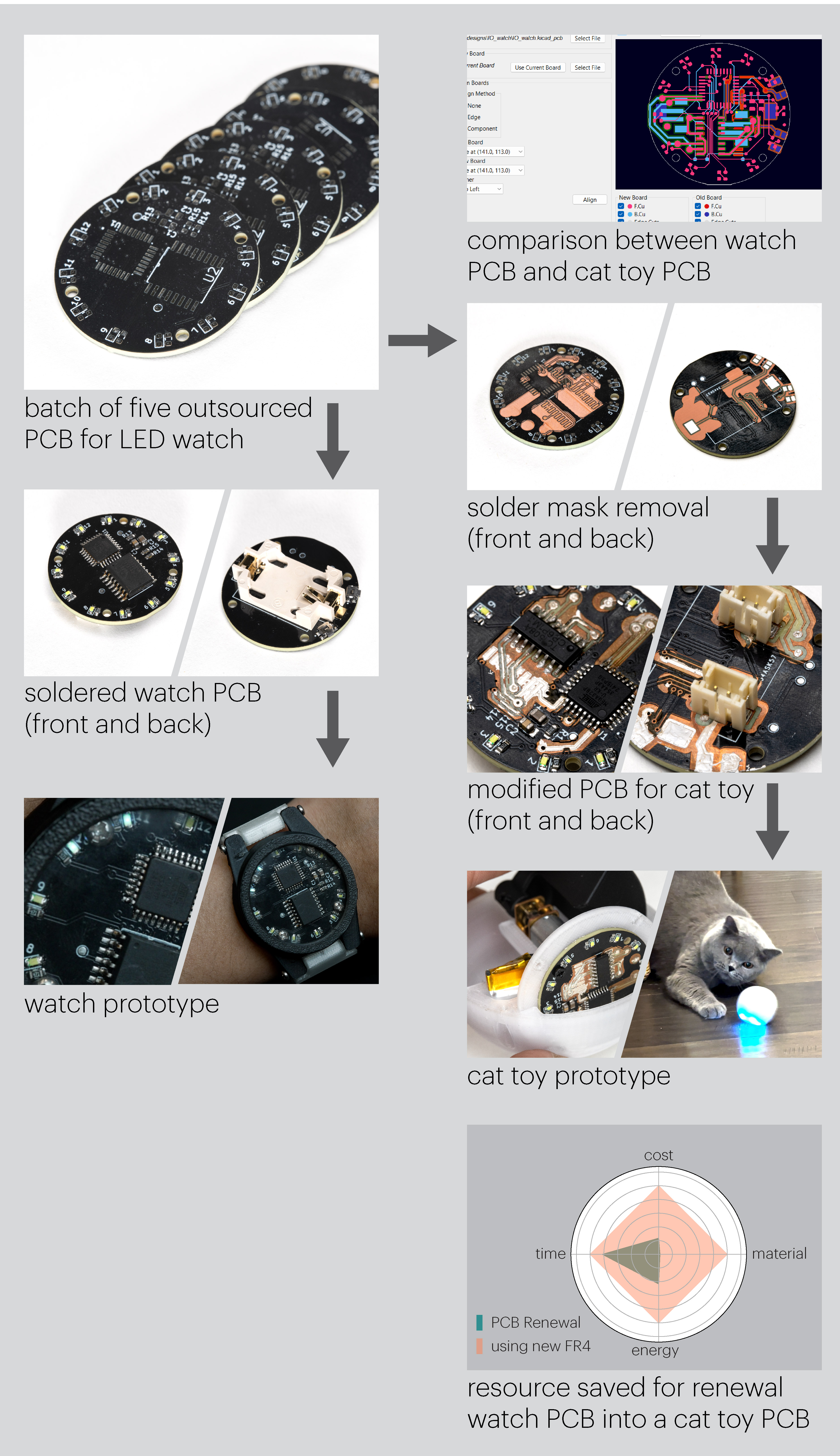}
  \caption{Renewal of an outsourced double-layer PCB: the left column shows the production of LED watch PCBs, including assembly into a working watch. The right column illustrates the renewal process, showcasing PCB design comparisons, solder mask removal, and board modifications for a cat toy, and its final assembly. A radar chart highlights the resources saved through \pcbrenewal versus creating a new FR4 board.}
  \Description{}
  \label{fig:addexamples}
\end{figure}

We began by outsourcing an open-source LED watch PCB \cite{IO_watch} to a small-batch manufacturer.
The PCB featured a standard double-layer configuration, a black solder mask, and a round shape (top image in the left column of Figure \ref{fig:addexamples}). 
Since the manufacturer requires a minimum order quantity of five PCBs, we had several extra boards remaining after successfully assembling the LED watch (shown in the bottom images of the left column in Figure \ref{fig:addexamples}). 
Typically, such boards are difficult to reuse in other projects due to their specific design. 
However, with the \pcbrenewal approach, these surplus PCBs can be easily repurposed. In this case, they were modified to function as the controller for an interactive cat toy ball.

The renewal process for an outsourced, double-sided PCB is largely identical to that of an in-house, CNC-milled PCB, with two exceptions: the removal of the solder mask in the area to be modified, and the editing of vias, if necessary.

Removing the solder mask was based on the engraving profile generated from the KiCAD plug-in (top image in the right column of Figure \ref{fig:addexamples}). 
Specifically, the plug-in computed the areas of difference between the original LED watch PCB and the newly designed toy PCB for both sides. These differential areas were then sent to a $CO_2$ laser cutter, which selectively removed the solder mask and exposed the copper conductors using rastering mode (row two in the right column of Figure \ref{fig:addexamples}). 
Alternatively, the solder mask can be removed using either a \SI{1064}{\nano\meter} wavelength fiber laser \cite{10.1145/3526113.3545652} or a diode laser \cite{RAELE2017475}.

In the new toy ball circuit design, new trace areas required electrical connections between both sides of the PCB, while some existing vias from the original PCB needed to be removed. To achieve this, undesired vias were drilled out using a square end mill during the engraving process. 
New vias are created using the same process, followed by either manual soldering into these through-holes to establish electrical connections or filling the entire via hole with conductive epoxy. 
The modified PCB is shown in row three of the right column in Figure \ref{fig:addexamples}.

The repurosed PCB was then assembled and installed into a custom 3D printed housing to complete the final cat toy (row four in the right column of Figure \ref{fig:addexamples}).
The renewal process significantly reduced material waste, manufacturing energy, and fabrication time. 
The sustainability modeling results are presented at the bottom of right column in Figure \ref{fig:addexamples}.
These estimates assume that the benchmark board is made using a CNC FR-4 board and account for the energy required for laser cutting during solder mask removal.

\pcbrenewal notably provides a much shorter turnaround time compared to ordering new PCBs from a manufacturer. Furthermore, it eliminates shipping-related energy costs, making \pcbrenewal a more efficient and sustainable solution.

\section{Discussion}
\pcbrenewal enables multiple iterations on a single FR-4 substrate, both within and across projects, promoting more sustainable PCB making practices. However, this approach also has its limitations. 
In this section, we discuss these limitations and outline potential future research opportunities.

\subsection{Unpacking Sustainability Benefits and Trade-Offs}
Across various examples and design iterations, we observed consistent savings in materials, costs, and energy, though time savings varied. 
For example, in the iteration of the camera roller for the same project, \pcbrenewal saved up to 60\% of the time by re-engraving only a small section of copper rather than engraving all traces on a fresh substrate. 
In other cases, such as the ESPboy, \pcbrenewal required slightly more time than fabricating a new PCB due to the increased amount of editing required. 
From the timing perspective, if a circuit design is straightforward to mill, the renewal approach might not be time-efficient. This underscores that the decision between creating a PCB from scratch and using \pcbrenewal is case-dependent and dynamic. The sustainability model developed in Section~\ref{sus_eval}, along with its implementation in the software plug-in (Section ~\ref{software}), provides end-users a practical tool for making informed decisions by offering comprehensive comparison data for each design iteration.

However, our current sustainability model has its own limitations and can be further improved. 
For example, the time and energy costs associated with the delivery of outsourced PCBs are not currently factored in, even though delivery is often the most time-consuming aspect of the PCB manufacturing process. 
In fact, if delivery time is considered, renewing a factory-made PCB is almost always more time-efficient than ordering a new one.

Additionally, the current calculation of material savings is rudimentary, focusing solely on the total weight of the material involved. 
Ideally, the model would be more precise and informative if it considered the carbon footprint of the FR-4 material saved in comparison to the additional use of silver-epoxy. 
However, since carbon footprint data for silver-epoxy is unavailable, total weight remains one of the few standardized metrics accessible for comparing different materials. 
This limitation highlights the need for a more open-data approach to LCA~\cite{cirotha2019lca, martinez2016lca}, particularly as new materials are developed and introduced to the market.

\subsection{Automating \pcbrenewal}
While our current work has evaluated \pcbrenewal in terms of material, time, and energy costs, other practical factors must be considered, such as the increased likelihood of manual errors introduced during the renewal process. 
For example, manually depositing silver-epoxy may require skills and experience, while curing the epoxy-filled PCB necessitates transferring the board to an additional heating device. 
Additionally, cutting new traces on an existing board needs precise alignment, requiring users to carefully position the board in the CNC machine. 
For some users, these extra steps and the increased risk of manual mistakes are important trade-offs to consider when weighing \pcbrenewal against the simplicity of creating a new board from scratch.

We envision that a few simple upgrades to a desktop CNC machine could reduce some of the labor effort, making \pcbrenewal more accessible. 
For example, epoxy deposition could be automated with desktop CNC machines that support syringe extruders. 
In addition, the CNC cutting plate could be equipped with a heating element (e.g., a 3D printer heating bed), allowing the curing process to be integrated into the automated workflow within the CNC machine. 
Finally, alignment could also be automated, for example, through a camera-based calibration process. 
If these changes are implemented, we can potentially transform an off-the-shelf CNC machine designed for making PCBs into one that also supports the remaking or renewal process, promoting more sustainable PCB making practices.

\subsection{Supported PCB Materials and Types}
This paper focuses on the FR-4 PCB substrate, as it is the most commonly used material for both in-house and outsourced PCB production. However, other more environmentally friendly PCB substrates, such as paper-based FR-1 or cellulose-based FR-3, might also be compatible with the current workflow, though we have not explored these options. We suspect that working with FR-1 or FR-3 materials may require alternative conductive epoxies that cure through UV processes rather than heat, given these substrates' lower operating temperature. This suggests a potential future direction for comprehensively understanding the comparability of different  substrate materials and conductive epoxies.

Our method supports single- and double-layer PCBs, whether manufactured in-house or outsourced. While our example (Section~\ref{outsourced_pcb}) demonstrates the technical viability of renewing and updating externally manufactured PCBs, it depends on having access to the original circuit design. For commercial PCBs that are not open source, this requirement poses a significant obstacle.
To enable the renewal process for commercial PCBs, reverse engineering techniques utilizing X-ray tomography~\cite{botero2020automated, asadizanjani2017pcb} or computer vision~\cite{ProtoPCB} would be necessary. However, integrating this approach into the current workflow remains an open question and requires further research.

\subsection{Toward PCB Reuse in the Long Run}
Our work primarily explores the technical feasibility of \pcbrenewal. However, achieving a long-term impact in sustainable making requires understanding end-users' willingness to adopt \pcbrenewal, which necessitates deployment and active community engagement.

As a first step, we have open-sourced the \pcbrenewal software plug-in (Section~\ref{software}). 
Future deployment will allow us to explore integrating \pcbrenewal with other complementary methods that support PCB reuse. 
For example, the SolderlessPCB~\cite{10.1145/3613904.3642765} method enables the reuse of electronic components without soldering, while ecoEDA~\cite{10.1145/3586183.3606745} facilitates  component reuse across multiple projects. 
It would be interesting to explore whether a more integrated and comprehensive PCB reuse system could influence end-users' PCB making and usage practice over time.

Finally, while this paper primarily considers \pcbrenewal in the context of individual PCB fabrication, it also holds potential for industrial-scale recycling.
For example, integrating a PCB layout recognition system into recycling facilities could potentially enable centralized operations to adopt \pcbrenewal, allowing useful PCBs to be repurposed before entering the waste stream.
Investigating industrial applications could uncover new opportunities for \pcbrenewal on a larger scale.

\section{Conclusion}
In this paper, we introduced \pcbrenewal, a novel technique that ``erases'' and ``reconfigures'' existing circuit traces.
We presented \pcbrenewal workflow and evaluate its electrical performance and mechanical durability. 
We modeled the sustainability impact of \pcbrenewal by calculating the material usage, cost, power, and time consumption for renewing PCB versus using new substrates. 
We implemented a custom EDA software plug-in that guides epoxy deposition, generates updated profiles, and calculates resource use.
We showcased the effectiveness of \pcbrenewal with a set of walkthrough examples, and concluded the paper by discussing its limitations and proposing future directions.

\begin{acks}
We thank Sandbox, the Jagdeep Singh Family Makerspace, for providing access to tools during the development and documentation of this project. We also thank our cat, Zhu Yuanzhang, for enjoying the custom cat toy and serving as a great model for Figure~\ref{fig:addexamples}. An LLM service was used exclusively for proofreading.
\end{acks}
\bibliographystyle{ACM-Reference-Format}
\bibliography{sample-base}
\end{document}